\documentclass[useAMS,usenatbib,usegraphicx]{mn2e}
\usepackage{bm,amssymb}
\usepackage{amsmath}
\usepackage{times}
\usepackage{rotating}
\newcommand{\bc}{\begin{center}}
\newcommand{\ec}{\end{center}}
\newcommand{\be}{\begin{equation}}
\newcommand{\ee}{\end{equation}}
\newcommand{\ba}{\begin{array}}
\newcommand{\ea}{\end{array}}
\newcommand{\bea}{\begin{eqnarray}}
\newcommand{\eea}{\end{eqnarray}}
\def\ga{\mathrel{\mathchoice {\vcenter{\offinterlineskip\halign{\hfil
$\displaystyle##$\hfil\cr>\cr\sim\cr}}}
{\vcenter{\offinterlineskip\halign{\hfil$\textstyle##$\hfil\cr
>\cr\sim\cr}}}
{\vcenter{\offinterlineskip\halign{\hfil$\scriptstyle##$\hfil\cr
>\cr\sim\cr}}}
{\vcenter{\offinterlineskip\halign{\hfil$\scriptscriptstyle##$\hfil\cr
>\cr\sim\cr}}}}}
\def\la{\mathrel{\mathchoice {\vcenter{\offinterlineskip\halign{\hfil
$\displaystyle##$\hfil\cr<\cr\sim\cr}}}
{\vcenter{\offinterlineskip\halign{\hfil$\textstyle##$\hfil\cr
<\cr\sim\cr}}}
{\vcenter{\offinterlineskip\halign{\hfil$\scriptstyle##$\hfil\cr
<\cr\sim\cr}}}
{\vcenter{\offinterlineskip\halign{\hfil$\scriptscriptstyle##$\hfil\cr
<\cr\sim\cr}}}}}
\def\mkm{{\mu}{\rm m}}

\def\is{interstellar }
\def\degr{\hbox{$^\circ$}}

\def\vec#1{\ensuremath{\mathchoice{\mbox{\boldmath$\displaystyle#1$}}
{\mbox{\boldmath$\textstyle#1$}}
{\mbox{\boldmath$\scriptstyle#1$}}
{\mbox{\boldmath$\scriptscriptstyle#1$}}}}

\title[Polarizing efficiency and grain growth]
{Polarizing efficiency as a guide of grain growth and
interstellar magnetic field properties}
\author[Voshchinnikov, Il'in, Das]{N.V. Voshchinnikov$^1$\thanks{E-mail:
    n$_{.}$voshchinnikov@spbu.ru}, V.B. Il'in$^{1,2,3}$
and H.K. Das$^{4}$ \\
$^1$Sobolev Astronomical Institute, St. Petersburg University,
Universitetskii prosp., 28, St. Petersburg 198504, Russia\\
$^{2}$Main (Pulkovo) Astronomical Observatory, 
 Pulkovskoe sh. 65, St. Petersburg 196140, Russia \\
$^{3}$St.~Petersburg State University of Aerospace Instrumentation,
Bol. Morskaya 67, St. Petersburg 190000, Russia \\
$^{4}$IUCAA, Post Bag 4, Ganeshkhind, Pune 411 007, India}

\date{Accepted 2016 July 18; Received 2016 July 15; in original form 2015 November 21}

\pagerange{\pageref{firstpage}--\pageref{lastpage}} \pubyear{2016}

\begin{document}
\label{firstpage}
\maketitle

\begin{abstract}
We interpret the relation between the
polarizing efficiency $P_{\max}/E(B-V)$  and the wavelength of
the maximum polarization $\lambda_{\max}$ observed for 17 objects
(including 243 stars) separated into two groups:
``dark clouds'' and ``open clusters''.
 The objects are assigned to one of the groups according to
the distribution of the parameter $\lambda_{\max}$.
 We use the model of homogeneous silicate and carbonaceous
 spheroidal particles  with the
imperfect alignment and  a time-evolving size distribution.
The  polarization is assumed to be  mainly produced by large
silicate particles with the sizes $r_{V} \ga r_{V,\rm cut}$.
The models with the initial size distribution
reproducing the average curve of the interstellar extinction
fail to explain the values of $\lambda_{\max} \ga 0.65\,\mkm$
observed for several dark clouds.
 We assume that the grain size distribution is modified due to accretion
and coagulation, according to the model of Hirashita \& Voshchinnikov (2014).
After including the evolutionary effects, $\lambda_{\max}$
shifts to  longer wavelengths on time-scales
$\sim 20 (n_\mathrm{H}/10^3 \mathrm{cm}^{-3})^{-1}$ Myr
where $n_\mathrm{H}$ is the hydrogen density in molecular clouds
where dust processing occurs.
 The ratio $P_{\max}/E(B-V)$ goes down dramatically
when the size of polarizing grains grows.
The variations of the degree and
direction of particle orientation influence this ratio only moderately.
We  have also found  that the aspect ratio of prolate grains
does not affect significantly the polarizing efficiency.
 For oblate particles, the shape effect is stronger but 
in most cases the polarization curves produced are too narrow
in comparison with the observed ones.
\end{abstract}


\section{Introduction}

Non-spherical dust grains are  responsible
for the interstellar polarization phenomenon.
The polarization becomes significant
when interstellar grains are well aligned and
have sizes comparable to the wavelength of the incident radiation.
 Alignment of dust particles can arise due to their
specific magnetic properties giving them the ability to efficiently
interact with the interstellar magnetic fields.

The \is linear polarization is characterized by the polarization degree
$P$ and the position angle $\theta_{\rm E}$ or $\theta_{\rm G}$ measured
either in the equatorial or galactic coordinate system, respectively.
 The polarization degree usually tends to
a maximum in the visual and gradually
decreases to the ultraviolet and infrared.
 The wavelength dependence of polarization $P(\lambda)$
is described by an empirical formula suggested in
\citet{serk73} and now called Serkowski curve
\be
P(\lambda)/P_{\max} = \exp [-K \ln^2 (\lambda_{\max}/\lambda)].
\label{serkk}
\ee
 This formula has three parameters: 
$P_{\max}$ is the maximum degree of polarization,
$\lambda_{\max}$ the wavelength corresponding to it, and
$K$ the coefficient  characterizing the width of the Serkowski curve.
The values of $P_{\max}$ in the diffuse interstellar medium usually
do not exceed 10\%, and the average
value of $\lambda_{\max}$ is 0.55\,$\mkm$
(\citealt{smf75}).
The parameter $K$ is related to
the half-width of the normalized  polarization curve
\be
W = \lambda_{\rm max}/\lambda_{-} -
 \lambda_{\rm max}/\lambda_{+},
\ee
where { \rm $\lambda _{-}, \lambda _{+}$ are such that }
$P(\lambda _{+}) = P(\lambda _{-}) = P_{\max }/2$ and
$\lambda _{-} < \lambda _{\max } < \lambda _{+}$.
The relation between $W$ and $K$ is as follows:
\be
W = \exp [(\ln 2/K)^{1/2}] - \exp [-(\ln 2/K)^{1/2}].
\label{wk}
\ee
Initially, the parameter $K$ was chosen to be 1.15 (\citealt{serk73}).
Later, \citet{wetal92} found
a dependence of $K$ on  $\lambda_{\max}$ in the Milky Way
\be
K = (1.66 \pm 0.09) \lambda_{\max} + (0.01 \pm 0.05),
\label{k92}
\ee
where $\lambda_{\max}$ is in microns.

\begin{table*}
 \centering
 \caption{Target objects.}\label{t_st1}
  \begin{tabular}{clcrrrcccc} \hline
$~N$ &~~~Object & $N_{\rm stars}$ & $l$~~ &  $\it b$~~ & $D$, pc
&  $\langle \lambda_{\max}  \rangle, \mu{\rm m} $
&  $\left\langle \frac{P_{\max}}{E(B-V)} \right\rangle $
& $\theta_{\rm 0, G} \pm \sigma_{\theta_{\rm 0, G}} $ 
& References \\
(1)&(2)&(3)&(4)&(5)&(6)&(7)&(8)&(9)&(10) \\
\hline
\multicolumn{10}{c}{Dark clouds} \\
~1 & Taurus: cloud~1     & 31 & 174 & --14 & 147 & 0.570 $\pm$ 0.049 &~6.01 $\pm$ 2.09           & 156.5 $\pm$ 13.1 & [1,2,3] \\
~2 & Chamaeleon~I        & 25 & 297 & --15 & 196 & 0.594 $\pm$ 0.064 &~9.46 $\pm$ 2.53           & 101.3 $\pm$ 10.8 & [4,5,6] \\
~3 & $\rho$~Oph: cloud~1 & 10 & 353 &  +18 & 133 & 0.684 $\pm$ 0.110 &~3.85 $\pm$ 1.56           &  31.4 $\pm$ 14.4 &  [7,8,9,10] \\
~4 & \phantom{$\rho$~Oph:} cloud~2 & 20 &     &      &     & 0.659 $\pm$ 0.059 &~6.19 $\pm$ 3.21 &  98.3 $\pm$ 11.2 &      \\
~5 & R~CrA: cloud~A1     &  9 & 359 & --18 & 148 & 0.758 $\pm$ 0.104 &~3.48 $\pm$ 1.62           & 132.0 $\pm$ 24.1 &  [4,5,11] \\
~6 & \phantom{R~CrA:} cloud~A2     & ~5 &     &      &     & 0.734 $\pm$ 0.080 &~2.21 $\pm$ 1.12 &  52.1 $\pm$ 14.3 &      \\
~7 & \phantom{R~CrA:} cloud~B      & ~5 &     &      &     & 0.690 $\pm$ 0.071 &~5.48 $\pm$ 2.57 & 153.4 $\pm$ 11.0 &      \\
~8 & CMa~R1: cloud~1     & ~5 & 225 & --1  & 690 & 0.637 $\pm$ 0.030 &~4.80 $\pm$ 2.06           & 132.4 $\pm$ 13.0 &  [12] \\
~9 & \phantom{CMa~R1:} cloud~2     & ~6 &     &      &     & 0.593 $\pm$ 0.149 &~2.31 $\pm$ 1.06 &  75.7 $\pm$ 24.3 &      \\
\noalign{\smallskip}
\multicolumn{10}{c}{Open clusters (and some dark clouds)} \\
10 & Taurus: cloud~2     & 13 & 174 & --14 & 147 & 0.556 $\pm$ 0.043 &~5.76  $\pm$ 1.61 &  26.1 $\pm$ 20.4 & [2,3] \\
11 & $\alpha$~Per        & 36 & 147 & --6  & 213 & 0.543 $\pm$ 0.051 &~8.10  $\pm$ 2.63 &  86.7 $\pm$ 13.8 & [13] \\
12 & Musca               & 16 & 300 & --9  & 171 & 0.577 $\pm$ 0.017 &12.62  $\pm$ 2.98 & 102.9 $\pm$  3.9 &  [4] \\
13 & Southern Coalsack   & 14 & 301 & --1  & 174 & 0.570 $\pm$ 0.040 &~6.31  $\pm$ 2.73 &  75.3 $\pm$ 17.4 &  [4] \\
14 & NGC~654             & ~6 & 129 & --0  &2410 & 0.555 $\pm$ 0.019 &~4.14  $\pm$ 0.71 &  87.3 $\pm$  4.5 &  [14] \\
15 & IC~1805             & 16 & 134 &  +1  &2400 & 0.540 $\pm$ 0.031 &~5.70  $\pm$ 1.14 &  96.1 $\pm$  5.0 &  [15] \\
16 & NGC~6124            & 11 & 341 &  +6  & 563 & 0.574 $\pm$ 0.039 &~3.40  $\pm$ 0.84 &  54.0 $\pm$  4.7 &  [16] \\
\noalign{\smallskip}
17 & Cyg~OB2             & 15 & ~80 &  +1  &1700 & 0.473 $\pm$ 0.073 &~2.07  $\pm$ 0.71 & 146.6 $\pm$ 26.2 &  [5,7] \\
\hline
\multicolumn{10}{l}{References:
 [1] -- \citealt{ef09};
 [2] -- \citealt{wetal01};
 [3] -- \citealt{hsu85};
 [4] -- \citealt{ap07};
 [5] -- \citealt{wetal92};
 }\\
\multicolumn{10}{l}{[6] -- \citealt{mcg94};
 [7] -- \citealt{maretal92};
 [8] -- \citealt{wilketal82};
 [9] -- \citealt{vrba93};
 [10] -- \citealt{Snow_2008};
 }\\
\multicolumn{10}{l}{[11] -- \citealt{vrba81}; 
 [12] -- \citealt{vrba87};
 [13] -- \citealt{ctv79};
 [14] -- \citealt{medhi08};
}\\
\multicolumn{10}{l}{[15] -- \citealt{medhi07};
 [16] -- \citealt{vergne10}.
}
\end{tabular}
\end{table*}

The ratio of $P_{\max}$ to the colour excess $E(B-V)$ or
{\rm visual} extinction $A_V$
is called the {\it polarizing efficiency}.
 There exists an empirical limit to  this ratio
(\citealt{smf75})
\begin{equation}
\frac{P_{\max}}{E(B-V)} 
\la 9 \%/{\rm mag.}
\label{pebv}
\end{equation}
or
\be
\frac{P_{\max}}{A_V} = \frac{P_{\max}}{R_V E(B-V)}
\la 3 \%/{\rm mag.}\,,
\label{pav}
\ee
where $R_V$ is the  total to selective extinction ratio.

A qualitative explanation of the relation~(\ref{k92}) between the width of
the \is polarization curve and the position of its maximum is connected
to the fact that
dust grains grow in the accretion and coagulation processes,
which leads to a narrowing of the particle size distribution
(see, e.g., \citealt{wetal92}).
A quantitative interpretation of the dependence
$K$($\lambda_{\rm max}$) was suggested by \citet{ag83}
who considered cylindrical particles and by \citet{voshchinnikov13}
and \citet{vh14} who used spheroidal grains.

However, all three parameters of the Serkowski curve have been determined
for a limited number of stars (less than 200; \citealt{vh14}).
 This is because one needs to perform
observations in more than four  or even five bands (\citealt{wetal92})
to find the width of the polarization
curve (parameter $K$) with a good accuracy.
In contrast, the parameters $P_{\max}$ and $\lambda_{\max}$ can be
determined with a sufficient accuracy  when stars are observed
just in three or four bands.
 Hence, there exists a significant number of lines of sight for which
both the polarizing efficiency and the wavelength of maximum
polarization have been estimated.

The polarizing efficiency
in a given direction depends on the size of polarizing and
non-polarizing grains, the particle shape and the degree and
direction of grain alignment.
In general, the dust grain characteristics and the magnetic field can vary
along the line of sight. As a basic model we use that
of imperfectly aligned spheroidal particles
in a regular magnetic field
that was earlier applied to simultaneously analyse the \is extinction
and polarization curves in a wide spectral range
(\citealt{vd08}; \citealt{dvi10}; \citealt{Siebenmorgen14}).
 We consider the time evolution of the grain size distribution
due to accretion and coagulation processes (\citealt{hv14}) and
involved new optical constants of grain materials (\citealt{jones12};
\citealt{jones13}).
 \citet[ hereafter VH14]{vh14} used such a model to investigate
the relation between $K$ and $\lambda_{\max}$ that are
the parameters of the Serkowski curve which characterize
its width and maximum {\rm position}, respectively.

In this paper, we focus on interpretation of the observed
dependencies of $P_{\max}/E(B-V)$ on  $\lambda_{\max}$
and show that the most important factor influencing
these dependencies {\it is the size of polarizing grains}.
The paper is organized as follows:
Sections~\ref{obs} and ~\ref{mod} give
a description of the observational data and the model used,
 Section~\ref{res} {\rm presents results of
our modelling  of the polarizing efficiency
and their discussion, and
Section~\ref{concl} contains the conclusions made.


\section{Observational data}\label{obs}

\begin{figure*}
\resizebox{8.5cm}{!}{\includegraphics{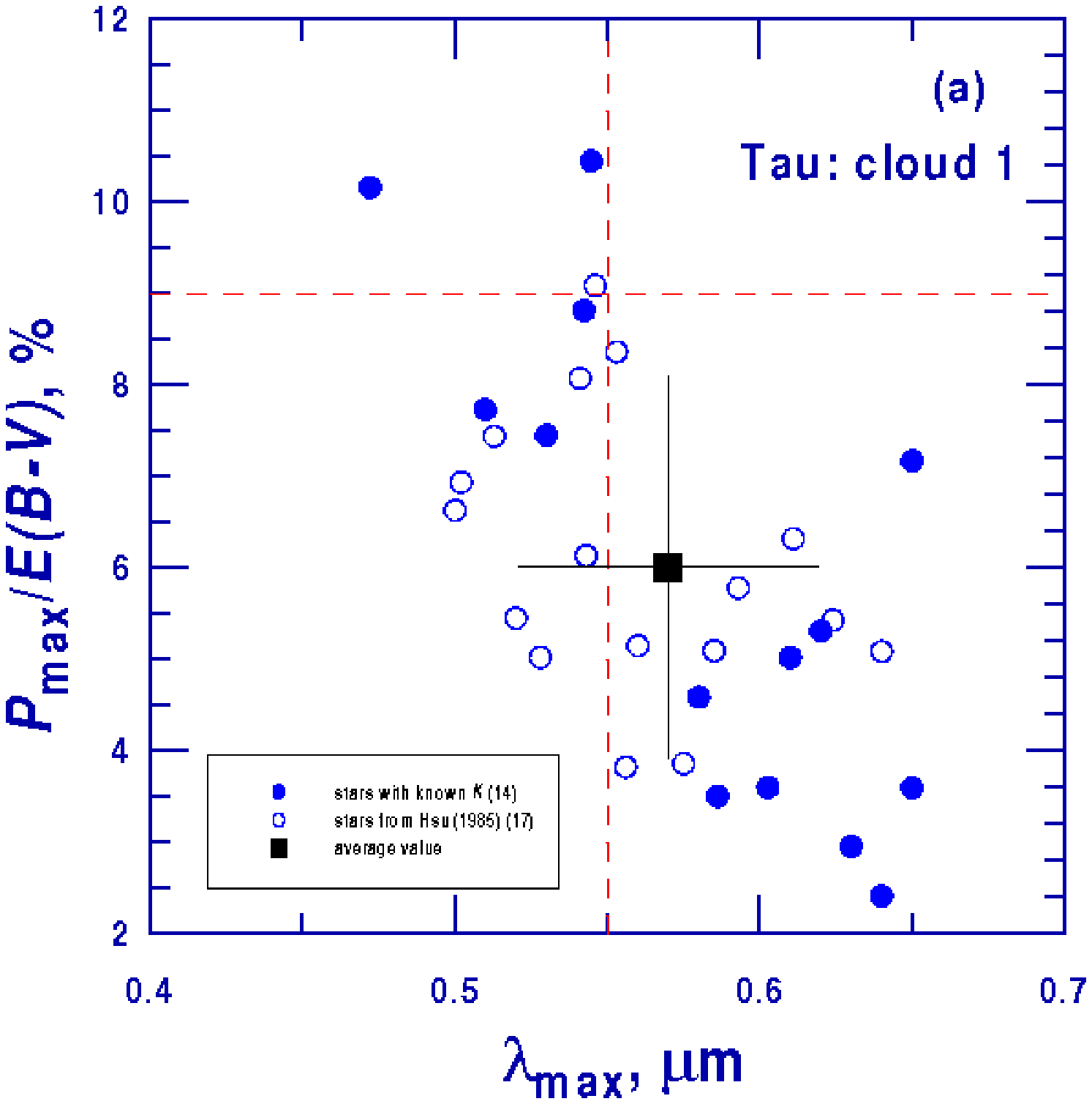}}
\resizebox{8.5cm}{!}{\includegraphics{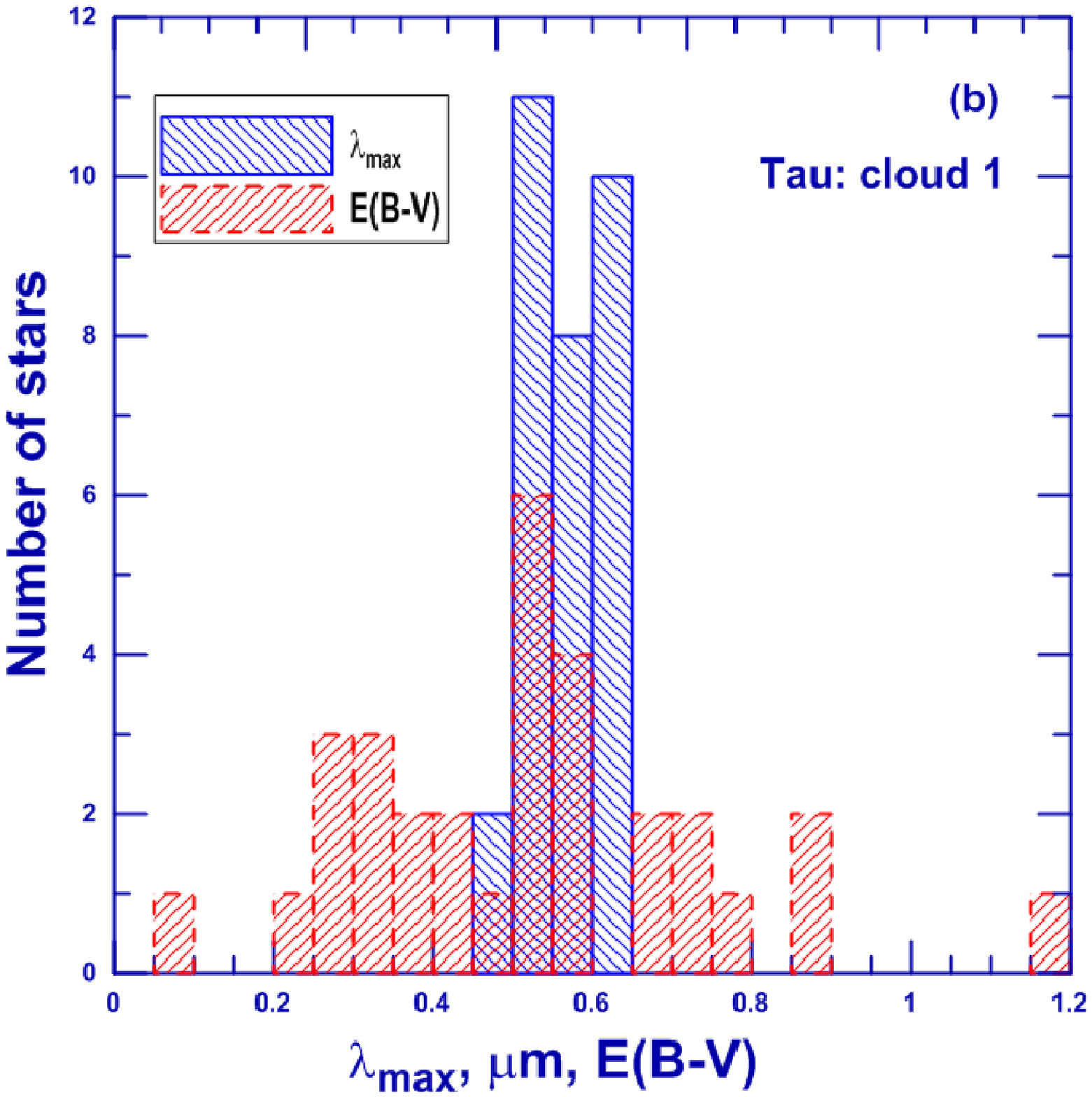}}
\resizebox{8.5cm}{!}{\includegraphics{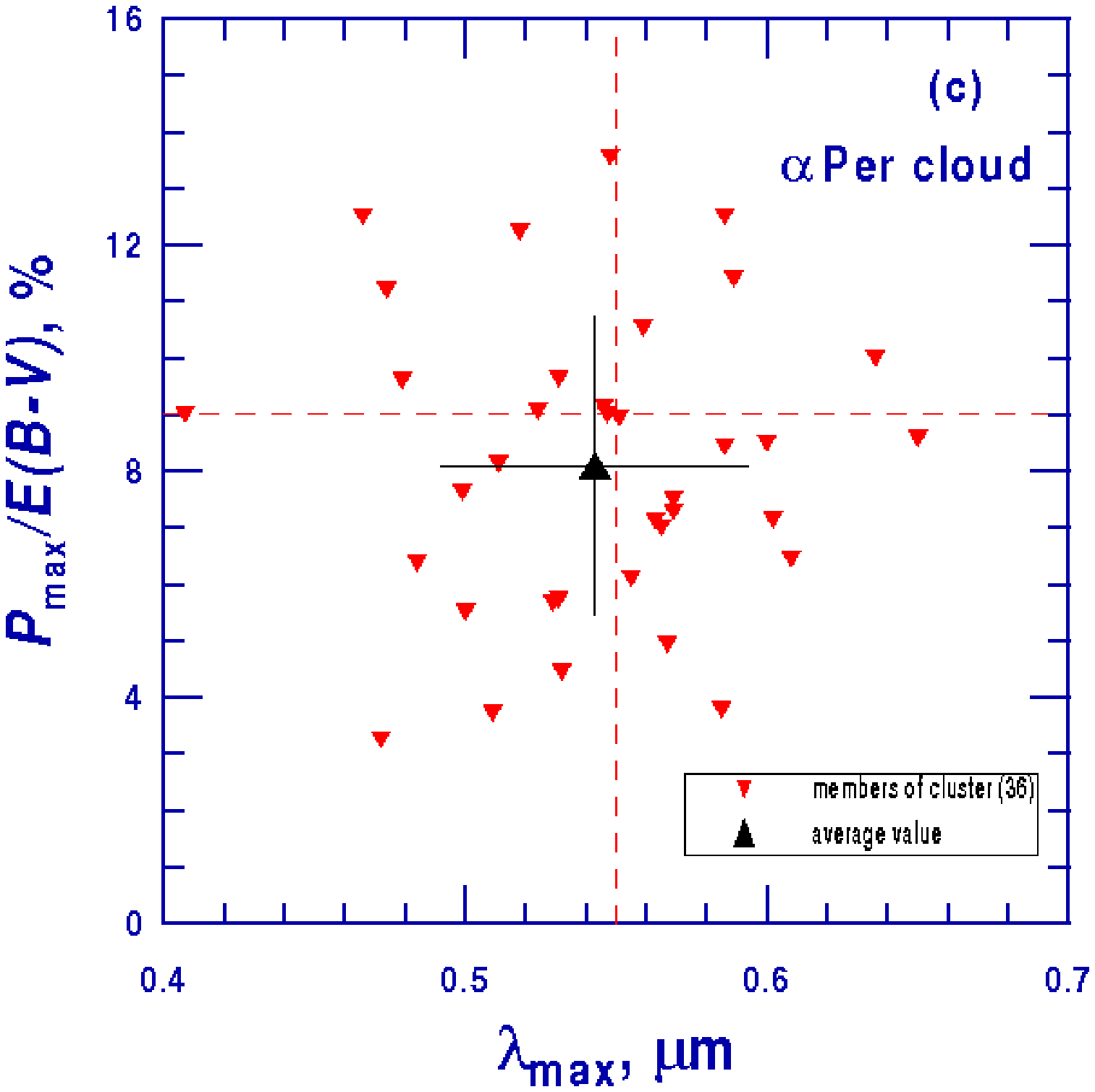}}
\resizebox{8.5cm}{!}{
\begin{sideways}
\includegraphics{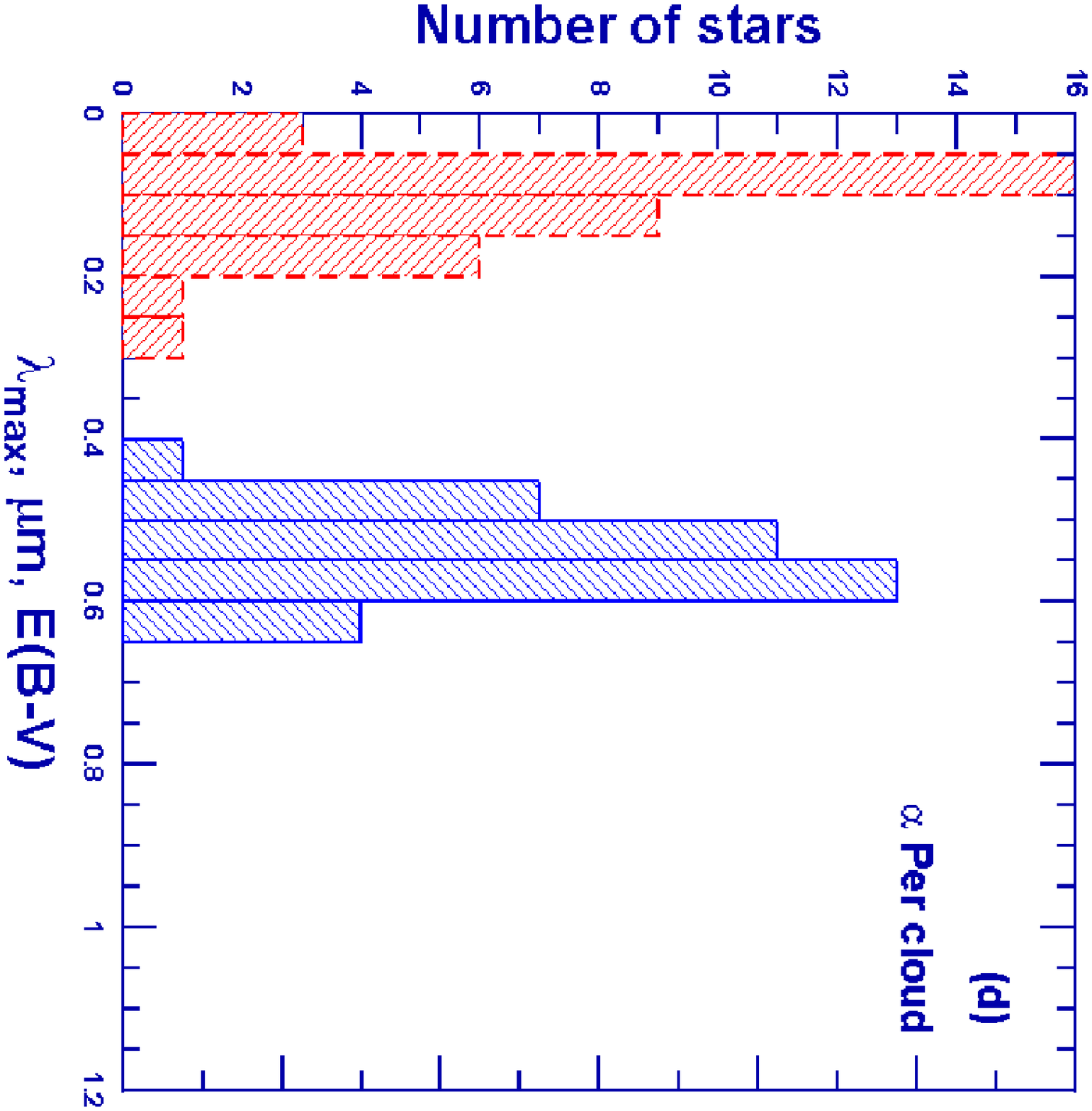}
\end{sideways}
}
\caption{Polarizing efficiency $P_{\max}/E(B-V)$
versus the  wavelength of maximum polarization $\lambda_{\max}$ for
dark cloud 1 in Taurus (a) and  cluster $\alpha$ Per (c).
The average values
 $\left\langle \frac{P_{\max}}{E(B-V)} \right\rangle $
are plotted with standard deviations.
Vertical and horizontal dashed lines
correspond to the mean \is value of $\lambda_{\max}=0.55\,\mu$m and
the upper limit of the polarizing efficiency given by equation~(\ref{pebv}).
Panels (b) and (d) show the histograms of the distributions of
$\lambda_{\max}$ and $E(B-V)$ for Taurus cloud 1 and
$\alpha$ Per cluster, respectively.}
\label{obs2}
\end{figure*}

We  have collected available observational data
on the  polarizing efficiency
$P_{\max}/E(B-V)$ and  the maximum polarization position
$\lambda_{\max}$ from the literature.
 We started with a list of 160 stars with  the known values of
the parameter $K$ considered in VH14.
From the list, we selected
stars closely located in the sky and with
similar values of the polarization position angle.
 We extended these groups by including stars with the known
values of $P_{\max}$, $\lambda_{\max}$, and $E(B-V)$
and excluded the stars with significant wavelength
rotation of the position angle and
large errors in observational data.
Morever, we chose only the probable
members of the stellar clusters.

Our final list contains 243 stars associated with 17 objects
that are either a dark cloud or a stellar cluster.
We separate the objects into two groups
according to the patterns seen on their
dependence of $P_{\max}/E(B-V)$ on  $\lambda_{\max}$ (see Fig.~\ref{obs2}).
All the objects are listed in Table~\ref{t_st1}
that includes the number of stars considered,
the coordinates and distances to the objects,
the average values of the wavelength of maximum polarization
with the deviations. The last column gives the sources of
the observational data.
 We took the distances to the objects from
\cite{knude} and the papers cited in Table~\ref{t_st1}.
Most of the objects are not very distant
and either belong to the Gould Belt or are close to it.
 The exceptions are NGC~654 and IC~1805 located in the
direction of the anti-center of the Galaxy where the \is medium
is transparent up to large distances (\citealt{lall14}).

\begin{figure}
\centerline{
\resizebox{\hsize}{!}{\includegraphics{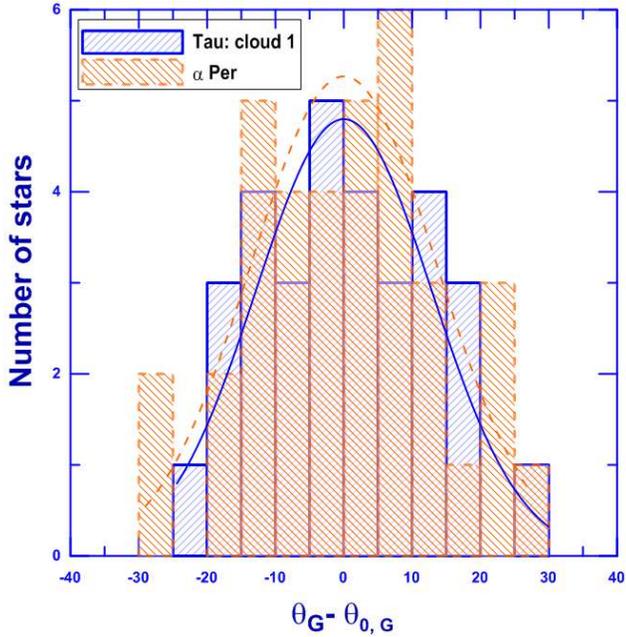}}
}
\caption{\rm Number distribution of the position
angles around the mean value $\theta_{\rm 0, G}$
for the dark cloud 1 in Taurus and cluster $\alpha$ Per.
Curves show the Gaussian fits of the observational data.}
\label{theta}
\end{figure}

For the clusters, polarization of their stars is mainly caused
by dust in the diffuse medium in the line of sight.
 On the plots we see a cloud of points around some mean
polarization efficiency, with $\lambda_{\max}$ being 
usually in the interval 0.5 -- 0.6 $\mu$m (see Fig.~\ref{obs2}c,d).
The clusters produce what we call the open cluster pattern.

For the dark clouds, when a part of the stars with measured
polarization are seen through outer layers of the cloud,
we meet another pattern called  the  dark cloud pattern.
 The stars observed mainly through 
diffuse medium, i.e. foreground, aside, distant stars, produce again
nearly the open cluster pattern, but the stars seen through the dark cloud
sometimes add a trend: the larger $\lambda_{\max}$, the smaller
the polarization efficiency (see Fig.~\ref{obs2}a,b).

Obviously, when the part of stars seen through the dark cloud 
is very small, we approach  the open cluster pattern,
 even though  the stars are at different distances,
while for the clusters  the distances are nearly the same.
This pattern is seen in our data for Taurus cloud 2, Musca cloud, and
Southern Coalsack. 

\begin{figure}
\resizebox{\hsize}{!}{\includegraphics{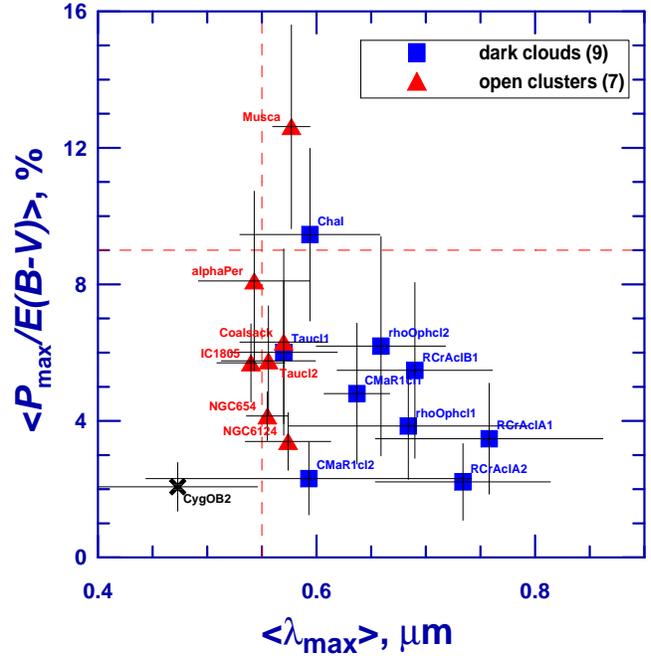}}
\caption{Average polarizing efficiency
$\langle P_{\max}/E(B-V) \rangle$
versus the average wavelength of maximum polarization
$\langle \lambda_{\max} \rangle$ for
dark clouds (squares) and open clusters (triangles)
in the Milky Way.
Vertical and horizontal dashed lines
correspond to the mean \is value of $\lambda_{\max}=0.55\,\mu$m and
the upper limit of the polarizing efficiency given by equation~(\ref{pebv}).
The number of stars, coordinates of objects, sources of data,
etc. are presented in Table~\ref{t_st1}.}
\label{obsd}
\end{figure}

The way we associate
the stars with the objects is to some extent  arbitrary
and is based on the data published
in the  papers listed in Table~1
(e.g., \citealt{vrba81} set off the cloud~A\footnote{\rm Besides this,
we distinguished clouds~A1 and A2.}
and cloud~B in the dark cloud R~CrA).
 In Taurus, two groups of stars with a relatively
uniform distribution of the position angles
(cloud 1, $\theta_{\rm G}=145\degr - 175\degr$ and
cloud 2, $\theta_{\rm G}=2\degr - 40\degr$)  were
described by \citet{voshchinnikov12} on the basis of the initial analysis
of \citet{mess97}.
 Stars in ``clouds'' are distributed around complexes of
molecular clouds and partly are projected on them.
 In these cases, the \is extinction can differ significantly,
 in contrast to stars of ``clusters''
when it lies in a rather narrow range
(cf. Figs.~\ref{obs2}b and ~\ref{obs2}d).
A negative correlation between
$P_{\max}/E(B-V)$ and $\lambda_{\max}$ for the  dark cloud 1 in Taurus
and a uniform distribution of data for $\alpha$ Per cluster
are well seen in the figures.
 Note that neither such a correlation nor its absence is
typical of the clouds and clusters.
Probably, the correlation is formed by local conditions and
is partly a result of observational selection.

Column (8) in Table~\ref{t_st1} shows
the average values of the polarizing efficiency
derived from observations.
These values do not take into account
the depolarization effect due to spatial variations
of the magnetic field $\vec{B}$.
This effect  can be found as the dispersion in
directions of polarization angle.
 A similar approach has been applied to optical polarization
(\citealt{mg91}; \citealt{jkd92}) and
submm polarization measured by {\it Planck}
(\citealt{planck32}; \citealt{planck35}; \citealt{planck44};
\citealt{guillet15}).

\citet{mg91} studied the spatial patterns and
distributions of the direction of the \is polarization for several
clouds, clusters and complexes of clouds.
 They found that the number distributions often had a single
local maximum with the dispersion of 0.2 -- 0.5 radians
and the clouds had a larger dispersion than the clusters.
 We {\rm calculate} the mean value of the position angle
$\theta_{\rm 0, G}$ and its dispersion $\sigma_{\theta_{\rm 0, G}}$
for our objects (see column 9 in Table~\ref{t_st1}).
 Figure~\ref{theta} presents the number distribution of the position angle
for the dark cloud 1 in Taurus and cluster $\alpha$ Per together
with the Gaussian fits. 
 Note that the dispersions $\sigma_{\theta_{\rm 0, G}}$
are very close to the widths of Gaussian curves
($13\degr1$ vs $12\degr9$ and $13\degr8$ vs $13\degr6$
for dark cloud 1 in Taurus and cluster $\alpha$ Per, respectively).
Though we used the data different from those of \citet{mg91},
the dispersions found for $\alpha$ Per and Cha I do not differ more
than by 20\%. For the regions in Ophiuchus and Taurus the difference
is about 50\% mainly because we considered separate clouds, while
\citet{mg91} large complexes.

 Figure~\ref{obsd} shows the average polarizing efficiency
versus the average wavelength of maximum polarization for target objects.
One clearly sees} two patterns in distribution of the
$\langle \lambda_{\max} \rangle$ values:
 for the open clusters and some clouds, the mean values of $\lambda_{\max}$
are grouped around the average \is value equal to 0.55 $\mu$m with
rather small deviations, while for most dark clouds,
the values of $\langle \lambda_{\max} \rangle $ may be significantly
larger than 0.55 $\mu$m and have a  wider scattering.
Note also that in two cases (Cha~I and Musca) the average
polarizing efficiency exceeds the upper observational limit
given by equation~(\ref{pebv}).

We singled out the association Cyg~OB2 which
still remains one of the most actively
studied objects in the Galaxy (see a history of its investigations in
\citealt{chentsov13}).
 The polarimetric studies of stars in Cyg~OB2 have demonstrated
a low polarizing efficiency and unusually small values of the maximum
polarization wavelength.
 Therefore, Cyg~OB2, marked by a cross in Fig.~\ref{obsd},
occupies an isolate position in the lower left corner.
 It is well known that in this direction the line
of sight is aligned along the spiral arm,
i.e. we should expect rather small polarization
because of a moderate  inclination of the magnetic field.
 From other side, the association Cyg~OB2 is rather distant.
So, depolarization of stellar radiation may occur due to
quite different polarization in  several clouds in the line of sight.
 Taking into account a contribution of
foreground polarization seems to lead to
an increase of the polarizing efficiency
(\citealt{mcm77}; \citealt{w15}).
 Evidently, polarization in Cyg~OB2 direction needs a further
analysis and discussion.

Undoubtedly, it would  be interesting to compare the optical polarization of
our objects with the submm polarization measured by {\it Planck}.
 Unfortunately, the published results were computed at $1\degr$ resolution
(\citealt{planck19}).
 So, the data can only suggest that this polarization is larger
in the region Chamaeleon--Musca than that in Ophiuchus, Taurus and R~CrA.


\section{Model}\label{mod}

Our model of  \is dust grains is exactly the same as
used by VH14.\footnote{\rm In previous modelling
(\citealt{dvi10}, \citealt{Siebenmorgen14}) we used other
optical constants and grain size distributions.}
We consider a mixture of silicate and carbonaceous homogeneous spheroids
 with certain distributions over sizes and orientations,
but a fixed ratio $a/b$, where $a$ and $b$ are the major and minor
spheroid semiaxes, respectively.
Both the silicate and carbonaceous particles contribute to extinction,
but the  \is polarization is assumed to be produced  mainly by
silicate particles.
 This assumption is supported, in particular,
by a correlation between the observed \is polarization degree and
the abundance of silicon in dust grains found in the work of \citet{vhpd12}.
We also suggest that small {silicate} particles {\rm are randomly
oriented when their sizes}  $r_V < r_{V,\rm cut}$, {\rm where}
$r_{V}$ is the radius of a sphere whose volume is equal
to that of the spheroid, $r_{V} = \sqrt[3]{a b^2}$ for prolate spheroids
and $r_{V} = \sqrt[3]{a^2 b}$ for oblate {\rm ones}.

Let us consider a dust cloud with the uniform magnetic field.
The angle between the line of sight and the magnetic field
is denoted by $\Omega$ ($0\degr \leq \Omega \leq 90\degr$).
The extinction and linear polarization of unpolarized stellar radiation
produced by aligned rotating spheroidal particles are
\bea
 A(\lambda)= 1.086 \sum_j \int\limits_0^D \int\limits_{r_{V,\min,j}}^{r_{V,\max,j}}
    \overline{C}_{{\rm ext},j}(m_{\lambda,j},a_j/b_j,r_{V},\lambda,\Omega)\,
 \nonumber \\ \times
    n_{j}(r_{V})\, dr_{V} \,dl\,,
\label{alam}
\eea
\bea
 P(\lambda)=  \int\limits_0^D \int\limits_{r_{V,{\rm cut},{\rm Si}}}^{r_{V,\max,{\rm Si}}}
    \overline{C}_{{\rm pol},{\rm Si}}(m_{\lambda,{\rm Si}},a_{\rm Si}/b_{\rm Si},r_{V},\lambda,\Omega)\,
 \nonumber \\ \times
    n_{{\rm Si}}(r_{V})\, dr_{V} \,dl \times 100\,\%\,,
\eea
where
\bea
 \overline{C}_{{\rm ext},j}
   = {\left ( \frac{2}{\pi}\right )^2}
   {\int\limits_{0}^{\pi/2}}{\int\limits_{0}^{\pi/2}}{\int\limits_{0}^{\pi/2}}
   \frac{1}{2}
   [C^{\rm TM}_{{\rm ext},j}(m_{\lambda,j},...,{\alpha})+C^{\rm TE}_{{\rm ext},j}] \,
 \nonumber \\ \times
    f_j(\xi, \beta, ...) \, d{\varphi}\, d{\omega}\, d{\beta} \,,
\label{cext}
\eea
\be
 \overline{C}_{{\rm pol},{\rm Si}}
   = {\frac{2}{\pi^2}}
   {\int\limits_{0}^{\pi/2}}{\int\limits_{0}^{\pi}}{\int\limits_{0}^{\pi/2}}
   {C}_{{\rm pol},{\rm Si}}\,
    f_{\rm Si}(\xi, \beta, ...) \, \cos 2{\psi} \, d{\varphi}\, d{\omega}\, d{\beta} \,,
\label{cpol}
\ee
and
\begin{equation}
{C}_{{\rm pol},{\rm Si}}=
   \frac{1}{2}
   [C^{\rm TM}_{{\rm ext},{\rm Si}}(m_{\lambda,\rm Si},...,{\alpha})-C^{\rm TE}_{{\rm ext},{\rm Si}}]
\notag\end{equation}
for prolate spheroids and
\begin{equation}
{C}_{{\rm pol},{\rm Si}}=
    C^{\rm TE}_{{\rm ext},{\rm Si}}(m_{\lambda,\rm Si},...,{\alpha})-C^{\rm TM}_{{\rm ext},{\rm Si}}
\notag\end{equation}
for oblate spheroids. \\
 Here $D$ is the distance to the star, $\lambda$ the wavelength,
$m_{\lambda,j}$, $a_j/b_j$ and $n_{j}(r_{V})$ are
the refractive index, aspect ratio and size distribution
of spheroidal particles of the $j$th kind
($j=$Si for silicate {\rm particles} and $j=$C for carbonaceous {\rm ones},
respectively),
${r_{V,\min,j}}$ and ${r_{V,\max,j}}$ are the minimum and maximum radii,
respectively,
$\alpha$ is the angle between  the wave vector of the incident radiation 
 and the rotation axis of a spheroid,
and $C^{\rm TM, \, TE}_{{\rm ext},j}$
the extinction cross-sections for two polarization modes
connected with the particle orientation relative to the electric vector of
the incident radiation (\citealt{bh83}).
 These cross-sections were calculated using a solution to the light
scattering problem for spheroids given by \citet{vf93}.
 The angle $\psi$ {\rm is} expressed through $\varphi, \omega, \beta, \Omega$
(see {\rm definitions of these} angles and relations between them,
e.g., in \citealt{dvi10}, \citealt{Siebenmorgen14}), and finally
${f}_j(\xi, \beta, ...)$ {\rm describes the} distribution of the particles
of the $j$th kind over orientations.

We calculate the extinction in the B, V  bands
as extinction $A(\lambda)$ (see equation~\ref{alam}) averaged over
 wavelengths within the corresponding passband as follows:
\setcounter{equation}{12}
\be
 A_X = - 2.5\, \log_{10}\, \int\limits_{\lambda_1}^{\lambda_2}
    {F}_X(\lambda)\, \exp \left [ \frac{A(\lambda)}{1.086} \right ]
     \,d\lambda\,,
\label{bvav}
\ee
where ${F}_X(\lambda)$ is the normalized transmittance curve for
a band X (=B, V) and $\lambda_1$ and $\lambda_2$ are the band
limits.

When selecting the dust materials, we followed VH14
and choose the amorphous silicate with a 10\% volume fraction
of Fe ('si\-li\-ca\-te$\_$FoFe10.RFI')
and hydrogen rich aliphatic carbon (a-C(:H) material
with a band gap $E_g=2.5$~eV).
The optical constants were taken from \citet{jones13} and
\citet{jones12} for the silicate and carbon, respectively.

An important constituent of the model is
the grain size distribution. It is obtained as a result
of  fitting  of the observed \is extinction and polarization
curves and may be a rather complicated function
(see \citealt{voshchinnikov12} for a review).
 Here we invoke the results of \citet{hirashita12} and \citet{hv14} who
investigated time evolution of the grain size distribution
caused by accretion and coagulation in an interstellar cloud.
 \citet{hv14} and VH14 examined whether dust grains
processed by these mechanisms can explain 
variations of the \is extinction and polarization curves
observed in the Milky Way.
 They assumed that an initial grain size distribution (for time $T = 0$)
should fit the mean Milky Way extinction curve (\citealt{weingartner01}).
 It was found that the observational data could be explained
provided the model is `tuned', i.e. when
coagulation of silicate dust is more efficient,
with the coagulation threshold being removed, and
coagulation of carbonaceous dust is less efficient 
compared to the original model.
 Thus, the time of grain processing $T$ is uniquely related to the
grain size distribution.  VH14 showed that  the time scale
$T \sim (30-50)(n_\mathrm{H}/10^3 \mathrm{cm}^{-3})^{-1}$ Myr,
the polarization maximum shifts to longer wavelengths
($\lambda_{\max}$ grows) and the polarization curve becomes
wider ($K$ decreases).
  A growth of $K$ and $\lambda_{\max}$  also occurs,
when we increase the cut-off size
$r_{V,\rm cut}$ that corresponds to the interface between
non-aligned and aligned grains.
 Both model parameters $T$ and $r_{V,\rm cut}$
influence the polarizing efficiency
$P_{\max}/E(B-V)$ ({\rm and} $P_{\max}/A_V$)  but
the degree and direction of grain alignment as well as the
particle type and shape may produce similar effects.

 Modelling requires a specification of the function
describing the distribution of particles over orientations 
according to a selected alignment mechanism. 
 Two mechanisms are most popular: 
the magnetic alignment based on the paramagnetic relaxation of grain material
containing about one percent of iron impurities
(DG mechanism; \citealt{dg51}), and  the 
radiative torque alignment (RAT alignment)
arising from an azimuthal asymmetry of the light scattering by non-spherical
particles (\citealt{dgs79}).
Both mechanisms encounter some difficulties.

The DG mechanism requires a stronger magnetic field than
 the  average galactic one, and the polarizing grains were assumed to
contain small clusters of iron (\citealt{js67}) and also needs
to be spun up to very high  velocities (\citealt{pur79}).

 The RAT mechanism has been updated rather recently 
(see \citealt{dw97} and the discussion in \citealt{alv15}).
 However, so far, it has been made rather 
at the qualitative level.  
 The theory of this mechanism is not based on careful
light scattering calculations of interstellar polarization
as the usage  of  the Rayleigh reduction factor 
 is appropriate just in the infrared part of spectrum
(see, e.g., \citealt{wetal08}).
 Note also the absence of  the correlation  between the submillimeter
polarization and dust temperature
predicted  by the RAT mechanism and not detected 
in {\it Planck} data (\citealt{planck11}; {\rm \citealt{guillet15}}).

Anyway, as the alignment function for the RAT mechanism 
has not been specified in enough detail
(\citealt{draine15}; \citealt{alv15}),
we utilized a properly modified function for 
the imperfect alignment of the spheroidal grains 
in Davis--Greenstein mechanism (IDG alignment).
 In this case, the distribution function ${f}^{\rm IDG}(\xi, \beta)$
depends on the orientation parameter $\xi$ and the angle $\beta$
 (\citealt{hg80}).
$\beta$ is the opening angle of the precession cone for
the particle angular momentum which precesses around the direction of
the magnetic field. 
 The alignment function is written as
\be
{f}^{\rm IDG}(\xi, \beta) = \frac{\xi \sin \beta}{(\xi^2 \cos^2 \beta + \sin^2 \beta)^{3/2}}.
\label{idg}
\ee
 The parameter $\xi$ depends on the particle size $r_V$,
the imaginary part of the magnetic susceptibility of a dust grain
$\chi''=\varkappa \omega_{\rm d} /T_{\rm d}$,
where $\omega_{\rm d}$ is the angular velocity of the particle,
hydrogen number density $n_{\rm H}$, magnetic field strength $B$, and
temperatures of dust $T_{\rm d}$ and gas $T_{\rm gas}$.
For particles of the $j$-th kind,
\be
\xi^2_j = \frac{r_V +\delta_{0,\,j}^{\rm IDG} (T_{{\rm d},j}/T_{\rm gas})}{r_V
+ \delta_{0,\,j}^{\rm IDG}},
\label{xi} \ee
where
\be
\delta_{0,\,j}^{\rm IDG} = 8.23\,10^{23} 
\frac{\varkappa_j B^2}{n_{\rm H} T_{\rm gas}^{1/2} T_{{\rm d},j}}\,\mkm
\label{delta} \ee
{\rm and we assume $T_{\rm gas} = 10 \, T_{\rm d}$.}
 In standard \is conditions, we have $\delta_{0} \approx 0.3 - 0.5\, \mu$m,
which allows one to account for the observed polarization (\citealt{dvi10}).
Note that the DG mechanism produces the ``right'' alignment when the
minor grain axes tend to align parallel to the magnetic field if
$T_{\rm d} < T_{\rm gas}$. Measurements show that this condition
is satisfied even in dense parts of \is clouds (\citealt{koumpia15}).

\begin{figure}
\centerline{
\resizebox{8.5cm}{!}{\includegraphics{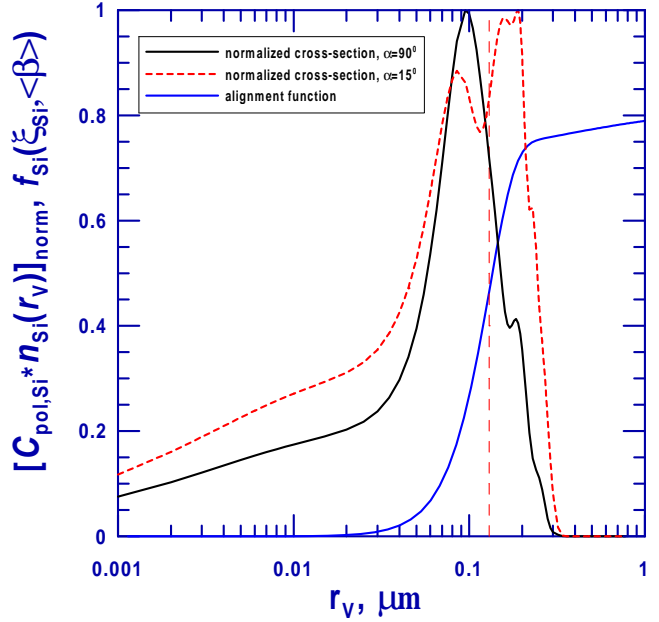}}
}
\caption{
Size  dependence  of the normalized polarization cross-sections for prolate
silicate spheroids with $a/b=3$ ($\lambda=0.55\,\mu$m)
and the alignment function $f_{\rm Si} (\xi_{\rm Si}, \beta)$
for $\beta$ equal to its average value $\langle \beta \rangle$ 
($T = 0$, $r_{V,\,\rm cut}=0.13\,\mu$m,
$\delta_{0,\,{\rm Si}}^{\rm IDG}=0.5\,\mkm$).
}
\label{align}
\end{figure}

We suggest the following modified IDG alignment
function with reduced alignment of small grains (see also \citealt{mathis86}):
\be
 {f}_j (\xi_j,\beta,...) =
\left [  1- \exp (-r_V/r_{V,{\rm cut},j})^3 \right ] \, \times \,
 {f}^{\rm IDG}_j(\xi_j, \beta) \,,
\label{fj}
\ee
where $r_{V,{\rm cut},j}$ is a cut-off parameter.
The alignment function (\ref{fj}) gives a smooth
switch from non-aligned to aligned grains.
It is shown in Fig.~\ref{align} for the model with the
parameters: $T=0$ Myr, $r_{V,\,\rm cut}=0.13\,\mu$m and
$\delta_{0,\,{\rm Si}}^{\rm IDG}=0.5\,\mkm$ (see Fig.~\ref{p0}).
 The alignment function was calculated for several grain sizes
as a function of the  mean precession cone angle $\langle \beta \rangle$
as given by equation~(5) in \citet{ag83}.
{\rm It should be emphasized that our}
function ${f}_j(...)$ qualitatively agrees with that expected
for the RAT mechanism.

Figure~\ref{align}  also shows  
the size dependence of the product  
${C}_{{\rm pol},{\rm Si}} (r_{V},\alpha,...)\,n_{{\rm Si}}(r_{V})$ 
for $\lambda=0.55\,\mu$m  normalized to its  maximum value.
 We consider the non-rotating prolate spheroids with $a/b=3$ in 
two orientations: $\alpha=90\degr$ and $15\degr$.
 The function $n_{{\rm Si}}(r_{V})$ corresponds to the initial grain size
distribution for time $T = 0$. It is clearly seen that
the contribution of larger particles to polarization
grows with  decreasing  the angle $\alpha$ 
between the wave vector and the particle symmetry axis.
The figure  demonstrates that a simplified
treatment of the partly aligned grains
using the Rayleigh reduction factor
may be a serious error.

 Our model assumes that the magnetic field
does not change its direction and strength in the line of sight to the star
(within the region where the polarization origins).
 However, turbulence  is known to affect the field.
A comparison of our model polarization $P$ with the observations
may require a correction reducing the polarization degree
as follows: $P_{\rm corrected} = F P$.
Now one often applies the so called depolarization factor $F$ that depends on
the ratio of the regular (uniform) component $\vec{B}_0$ of the magnetic
field to the total magnetic field
$\vec{B}_{\rm tot.}=\vec{B}_0+\vec{B}_{\rm t}$,
where $\vec{B}_{\rm t}$ is turbulent (random) component.
Thus, $F=1$ in the case of the regular magnetic field ($\vec{B}_{\rm t}=0$)
and $F=0.5$ in the case of the equipartition between the magnetic and
turbulent kinetic energy ($\vec{B}_{\rm t}=\vec{B}_0$).

\section{Results and discussion}\label{res}

We  have performed calculations  of the \is extinction
and polarization curves for prolate and oblate homogeneous spheroids
consisting of silicate and amorphous carbon.  The particles of 74
sizes in the range from $r_{V, \min}=0.001 \,\mkm$ to $r_{V, \max} = 1 \,\mkm$
with four aspect ratios $a/b= 1.5, 2, 3$ and 4 were utilized.
 The {\rm curves were} calculated for 77
wavelengths in the range  from $\lambda=0.2$ to $5\,\mkm$.
 We {\rm computed} the colour excess $E(B-V)$, parameter  $R_V$ and
parameters of the Serkowski curve $P_{\max}$,
$\lambda_{\max}$, and $K$.
 The response functions for the B and V bands
were taken from \citet{Straizys_1992}.

In this paper, we focus on  a relation between the
polarizing efficiency $P_{\max}/E(B-V)$ and $\lambda_{\max}$.
 The dependence of the width of the polarization curve (parameter $K$) on
the position of its maximum has been  analysed by VH14.
 It should be emphasized that $K$ and $\lambda_{\max}$
are mainly determined by the size distribution $n_{\rm Si}(r_{V})$
(depending on the time of grain processing $T$) and
the threshold on the size of aligned silicate grains
$r_{V,\rm cut}$. At the same time,  $K$ and $\lambda_{\max}$
are weakly  affected by the
degree ($\delta_{0}^{\rm IDG}$) and direction ($\Omega$) of the
particle orientation.
 However, it is not the case for the polarizing efficiency.
Therefore, we vary the alignment parameters 
$\delta_{0,\,{\rm Si}}^{\rm IDG}$ and $\Omega$
(keeping fixed the parameter $\delta_{0,\,{\rm C}}^{\rm IDG}=0.01\,\mkm$).

Our choice of $P_{\max}/E(B-V)$ as the polarizing efficiency
instead of $P_{\max}/A_V$ presents
the fact {\rm that} the ratio of the total to selective extinction $R_V$
given by observations is not reliable enough.
 Furthermore, grain processing in \is clouds leads to larger values
of $R_V$ (\citealt{hv14}; see also Table~\ref{t_ab}) that 
produces an additional trend in the behaviour
of the ratio $P_{\max}/A_V$ (see equation \ref{pav}) with a growth of $T$.

We start with a consideration of the  model
with prolate grains and $a/b=3$.
Variations of {\rm the} grain type and shape are
analysed in Sect.~\ref{ab_v}.

\subsection{Dust grains without processing ($T=0$ Myr)}

\begin{figure*}
\resizebox{8.5cm}{!}{\includegraphics{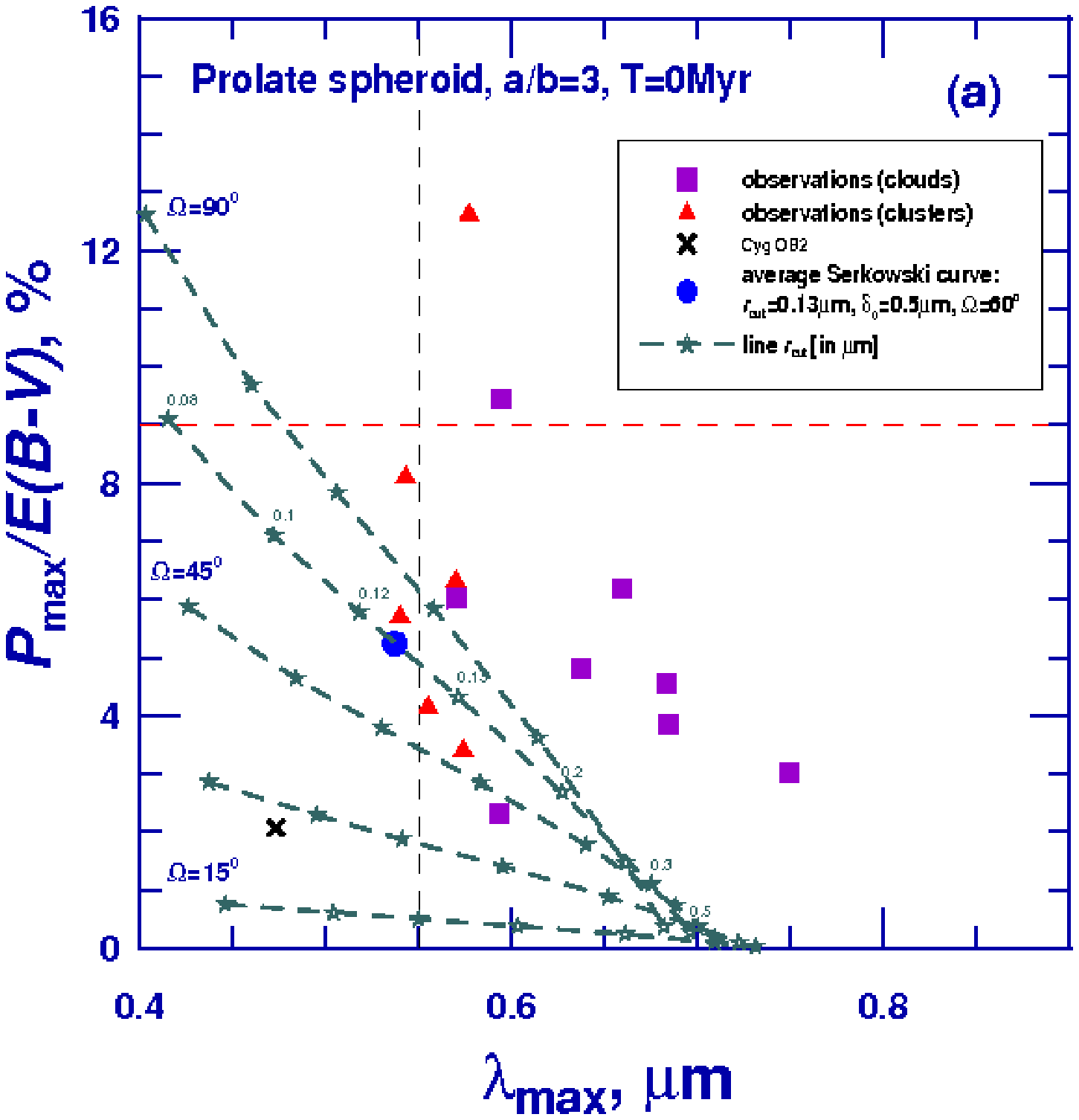}}
\resizebox{8.5cm}{!}{\includegraphics{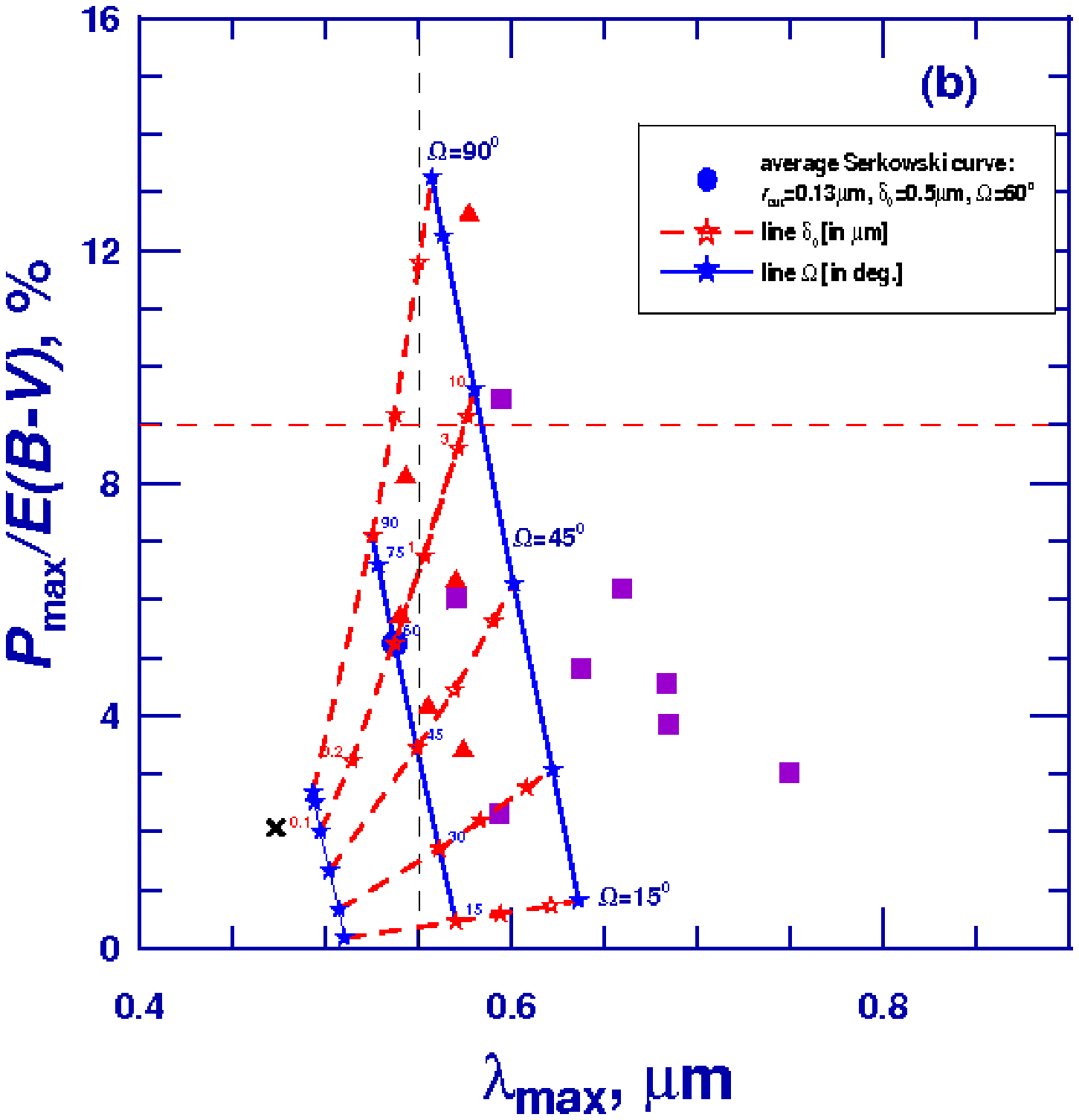}}
\caption{Polarizing efficiency $P_{\max}/E(B-V)$ versus the wavelength
of maximum polarization $\lambda_{\max}$.
 Symbols (filled squares and triangles and cross) show the average values
for clouds and clusters presented
in Table~\ref{t_st1} (columns 7 and 8) and Fig.~\ref{obsd}.
Vertical and horizontal dashed lines correspond to the mean \is
value of $\lambda_{\max}=0.55\,\mu$m and
the upper limit of the polarizing efficiency given by equation~(\ref{pebv}).
Stars ($\star$) connected with solid and dashed lines show theoretical
results calculated for the model: prolate spheroids, $a/b=3$.
Large filled circle corresponds to the model
with parameters: $T=0$ Myr, $r_{V,\,\rm cut}=0.13\,\mu$m,
$\delta_{0,\,{\rm Si}}^{\rm IDG}=0.5\,\mkm$, $\Omega=60\degr$
which is the closest to the average observational Serkowski curve
($\lambda_{\max}=0.55\,\mu$m, $K=0.92$).
Panels (a) and (b) illustrate the variations of the cut-off
parameter $r_{V,\,\rm cut}$ and the
degree ($\delta_{0,\,{\rm Si}}^{\rm IDG}$) and
direction ($\Omega$) of grain alignment, respectively.
 The depolarization factor $F$ can be involved by a proper vertical shift
of the theoretical curves.}
\label{p0}
\end{figure*}

First, we have calculated the data for diagrams
$P_{\max}/E(B-V)$ vs. $\lambda_{\max}$
for the initial size distribution of
silicate and carbonaceous  grains (Fig.~\ref{p0}).
 In this case, the average observational Serkowski curve
($\lambda_{\max}=0.55\,\mu$m, $K=0.92$) can be fitted
if $r_{V,\,\rm cut}=0.13\,\mu$m (VH14).
 If we assume $\delta_{0,\,{\rm Si}}^{\rm IDG}=0.5\,\mkm$,
$\Omega=60\degr$, {\rm and $F=1$,}
the polarizing efficiency of the \is medium
would achieve $P_{\max}/E(B-V)=5.23$\,\%/mag.
(large filled circle in Fig.~\ref{p0}).
 For $F<1$, this efficiency drops accordingly.

Figure~\ref{p0}a shows how variations of $r_{V,\,\rm cut}$ influence
the polarizing efficiency.  It is clearly seen that
$\lambda_{\max}$ grows and $P_{\max}/E(B-V)$ decreases
with a growing cut-off parameter,
i.e. for larger values of $r_{V,\,\rm cut}$,
the polarizing efficiency becomes  smaller
and the maximum of the curve $P(\lambda)$
shifts to longer wavelengths.
 This effect is more pronounced for larger values of $\Omega$.

An average size of polarizing dust grains
can be found using the following expression:
\be
\langle r_{V,{\rm pol,\,Si}} \rangle  = \frac{\displaystyle \int\limits_{r_{V,{\rm cut}}}^{r_{V,\max}} r_{V} n_{\rm Si}(r_{V}) \,{d}r_{V}}
{\displaystyle \int\limits_{r_{V,{\rm cut}}}^{r_{V,\max}} n_{\rm Si}(r_{V})\, { d}r_{V}}\,.
\label{rrr}\ee
 For the model which { \rm gives the best approximation }
 to the average observational Serkowski
curve ($r_{V,\,\rm cut}=0.13\,\mu$m),
$\langle r_{V,{\rm pol,\,Si}} \rangle = 0.17\,\mkm$.
 {\rm For}  $r_{V,\,\rm cut}$ {\rm increasing} from 0.08 to $0.30\,\mu$m,
$\lambda_{\max}$ grows, $P_{\max}/E(B-V)$ reduces  and
the average grain size $\langle r_{V,{\rm pol,\,Si}} \rangle$
increases from $0.12$ to $0.31\,\mkm$.

Note that for the models with $r_{V,\,\rm cut} \la 0.12 \,\mu$m
the polarization peaks in the blue part of spectrum
($\lambda_{\max}\la 0.5\,\mu$m) that is outside the range of 
the observational values of $\lambda_{\max}$ except for
the case of Cyg~OB2 {(Fig.~\ref{p0}a).}
The models with $r_{V,\,\rm cut} > 0.12 \,\mu$m can explain
only a small part of the observational data if we fix
the parameter $\delta_{0,\,{\rm Si}}^{\rm IDG}$.
 Also  it is evident that the cut-off parameter should
not exceed $\sim 0.2 - 0.3 \, \mkm$.

By increasing the degree of grain alignment and the angle $\Omega$
it is possible to reproduce almost all observations of
open clusters (Fig.~\ref{p0}b).
However, the models do not give polarization curves
with $\lambda_{\max} \ga 0.65\,\mkm$ observed in several dust clouds.
In this case, {\rm even very large values of
$\delta_{0,\,{\rm Si}}^{\rm IDG}$ would not do.}

Summarizing, we can conclude that the models with
the initial size distribution of {\rm unprocessed} dust grains
{\it fail} to explain the observational data with large values
of $\lambda_{\max}$.

\subsection{Processed dust grains ($T = 10 - 40$ Myr)}

\begin{figure*}
\resizebox{8.5cm}{!}{\includegraphics{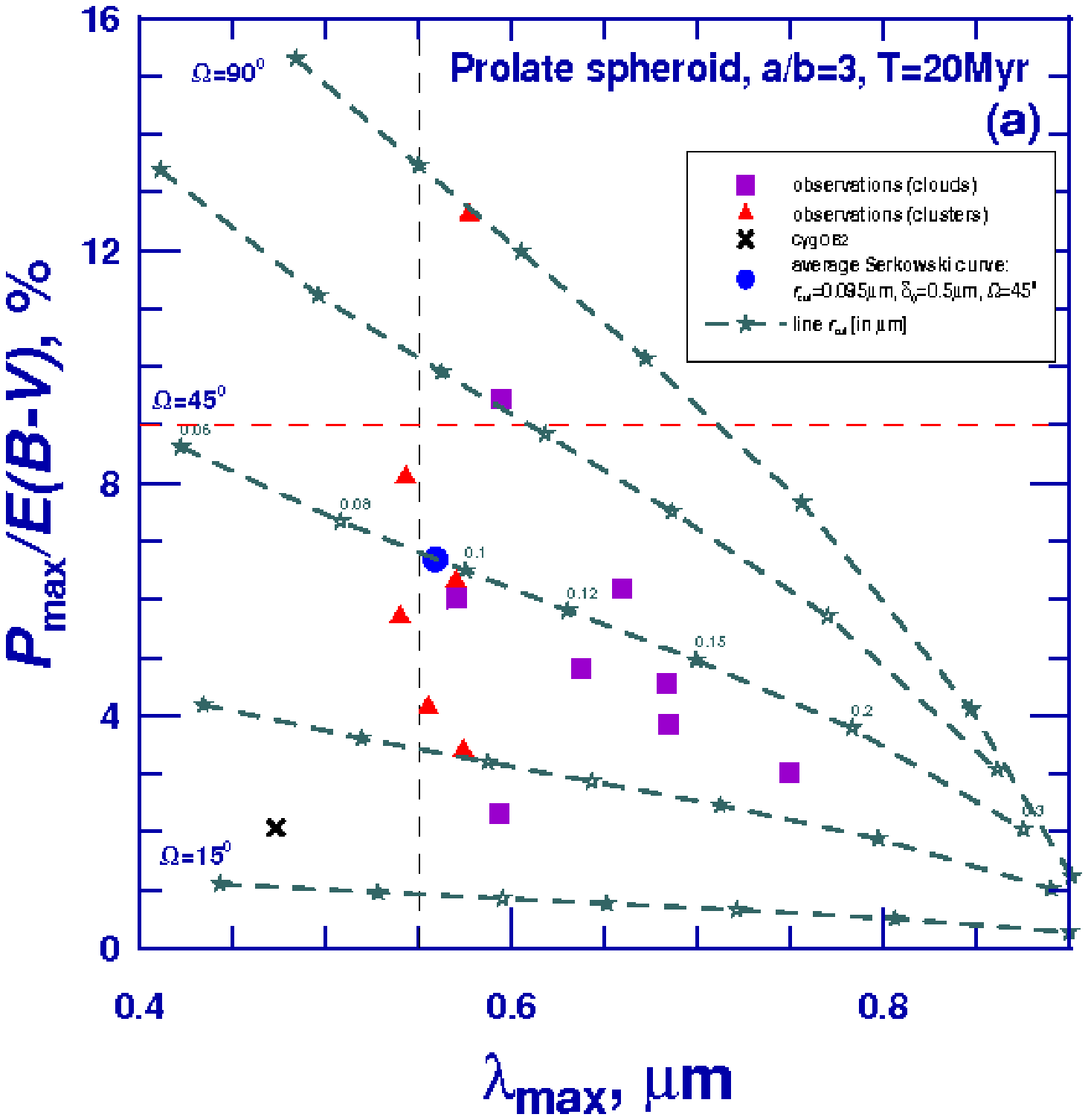}}
\resizebox{8.5cm}{!}{\includegraphics{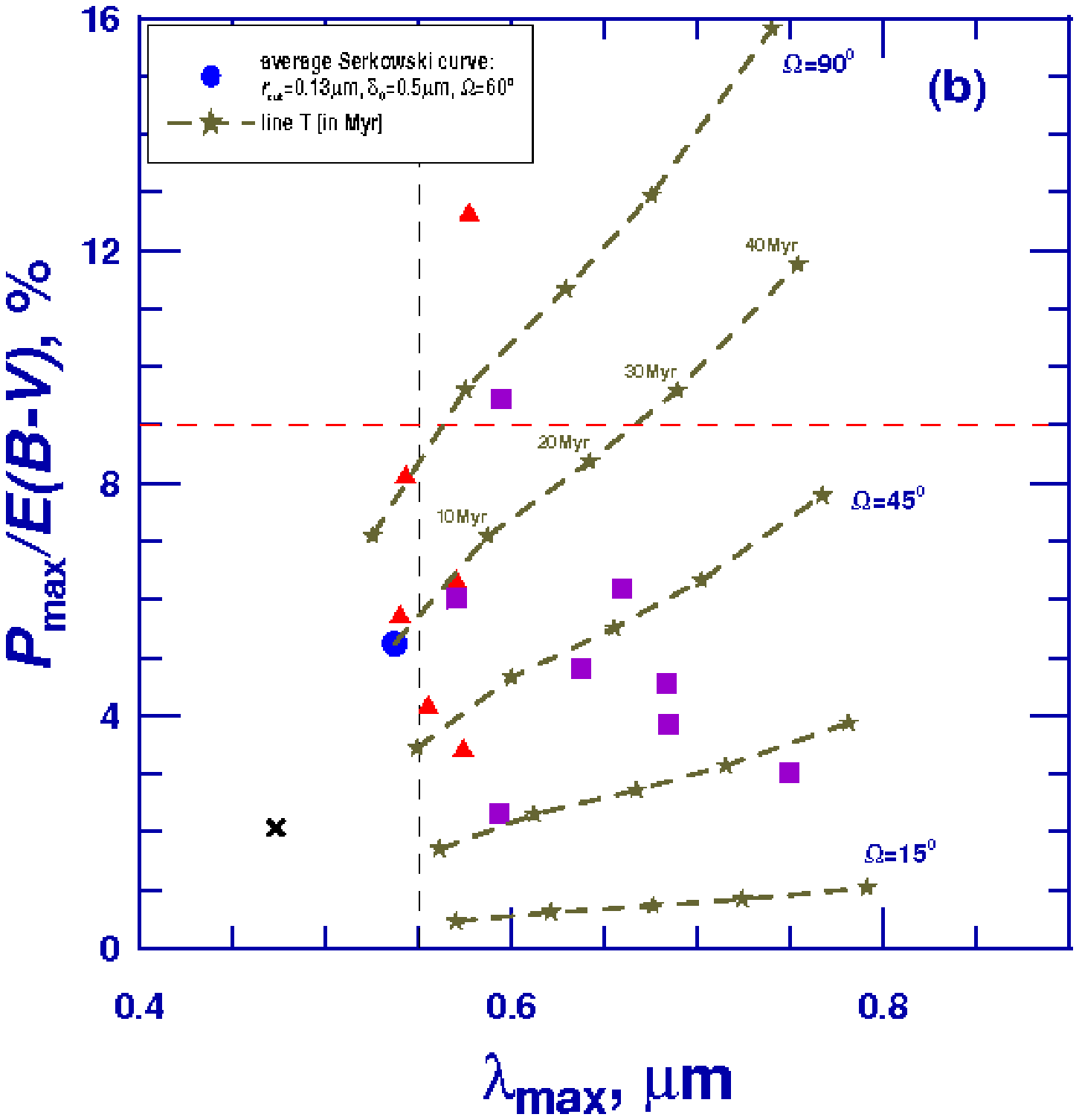}}
\caption{The same as in Fig.~\ref{p0} but now for the models with
different time of grain evolution $T$.
Panel (a) illustrates the variations the cut-off
parameter $r_{V,\,\rm cut}$
for the model with prolate spheroids, $a/b=3$,
$T=20$~Myr,
$\delta_{0,\,{\rm Si}}^{\rm IDG}=0.5\,\mkm$.
Large filled circle corresponds to the model
with $r_{V,\,\rm cut}=0.095\,\mu$m and $\Omega=45\degr$.
Panel (b) illustrates the variations of the time of grain evolution
for the model with parameters: $r_{V,\,\rm cut}=0.13\,\mu$m,
$\delta_{0,\,{\rm Si}}^{\rm IDG}=0.5\,\mkm$, $\Omega=60\degr$.}
\label{p20}
\end{figure*}

Next we investigate changes of the polarizing efficiency owing to
evolution of the grain size distribution.
 Our results are shown in Fig.~\ref{p20} for the
 tuned models (see \citealt{hv14}).
The left panel illustrates
the combined effects of $r_{V,\,\rm cut}$ and $\Omega$ variations
for the model with $T = 20$~Myr.\footnote{This
time of dust processing in a molecular cloud
 with the hydrogen density $n_\mathrm{H}=10^3$ cm$^{-3}$
(\citealt{hv14}).}
In this case, we have obtained
the polarization curve that is most close
to the average observational Serkowski one
when $r_{V,\,\rm cut}=0.095\,\mu$m,
$\langle r_{V,{\rm pol,\,Si}} \rangle = 0.15\,\mkm$ {\rm ($F=1$).}
 From a comparison of Figs.~\ref{p20}a and \ref{p0}a,
it is clearly seen that the evolutionary effects are very important:
 by varying $r_{V,\,\rm cut}$ and $\Omega$ in the model with $T = 20$~Myr
it is possible, in fact, to explain {\it all} observational data
both for open clusters and dark clouds {\rm when $F = 1$,
and all except for two largest values when $F=0.5$. }
 If we assume $\delta_{0,\,{\rm Si}}^{\rm IDG}=0.5\,\mkm$,
the models with the cut-off $0.08 \,\mkm  \la r_{V,\,\rm cut} \la 0.2\,\mkm$
and $\Omega \ga 30 \degr$ give the polarizing efficiency
and $\lambda_{\max}$ similar to the observed average values
 when $F<0.8$.

Note that an explanation of the available observational data
does not require the models with more processed dust grains ($T \ga 30$~Myr).
 This follows from Fig.~\ref{p20}b where we present the models with
increasing time of grain evolution and fixed other parameters.
 With growing $T$ both the polarizaing efficiency and the wavelength of
maximum polarization grow, and this effect is more pronounced if
the direction of the magnetic field is perpendicular
to the line of sight (i.e. for $\Omega=90\degr$).

\subsection{Models with prolate grains and $a/b=3$: summary}

Here we discuss the models in the context of simultaneous
variations of the four model parameters:
$r_{V,\rm cut, Si}$, $T$, $\delta_{0,\,{\rm Si}}^{\rm IDG}$ and
$\Omega$.  The first two parameters affect
the  size of polarizing and processed grains,
the last two parameters
determine the degree and direction of grain alignment.
 The type and shape of the particles (prolate spheroids with $a/b=3$)
remain the same in all the models.

\begin{figure*}
\resizebox{8.5cm}{!}{\includegraphics{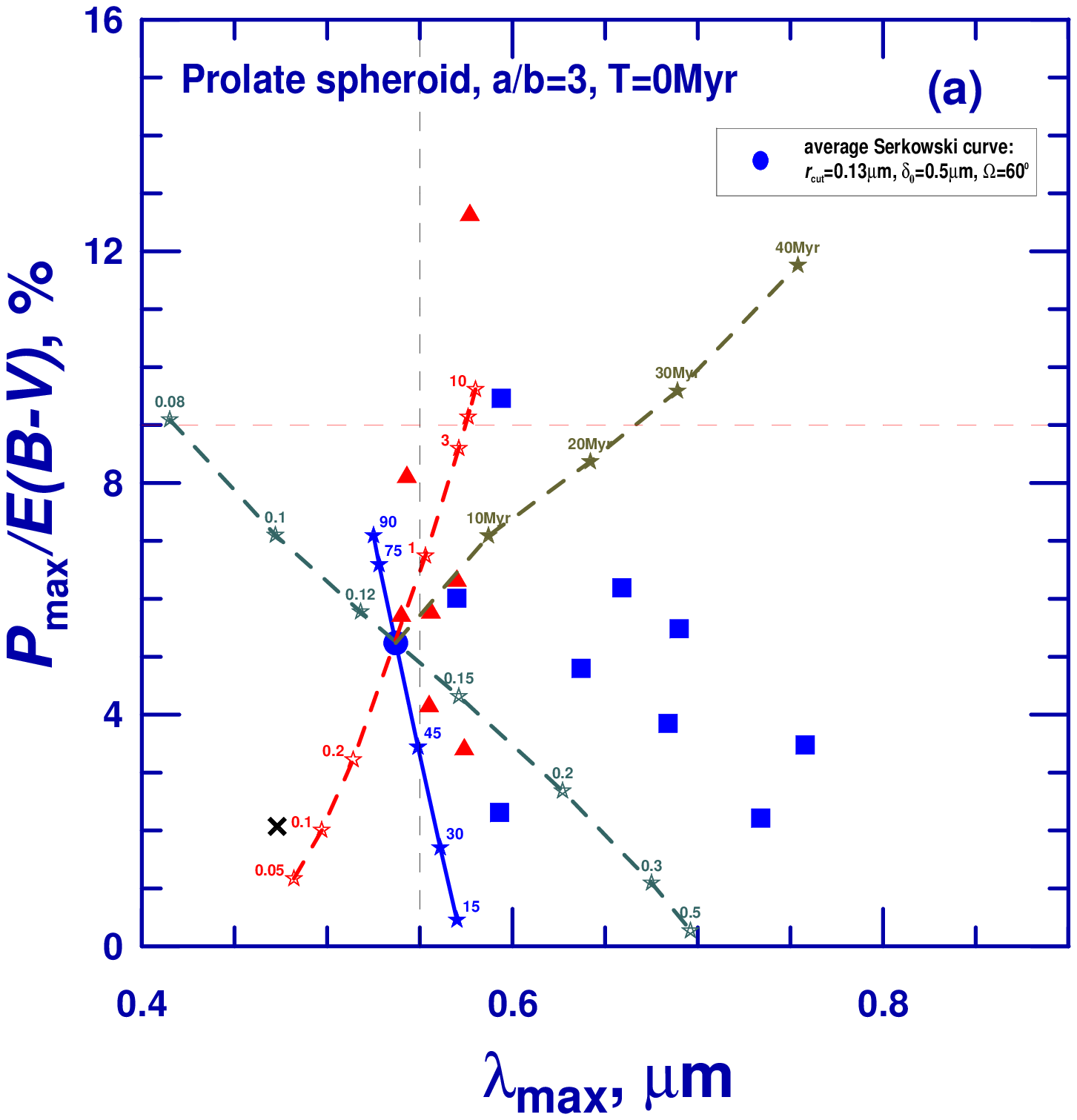}}
\resizebox{8.5cm}{!}{\includegraphics{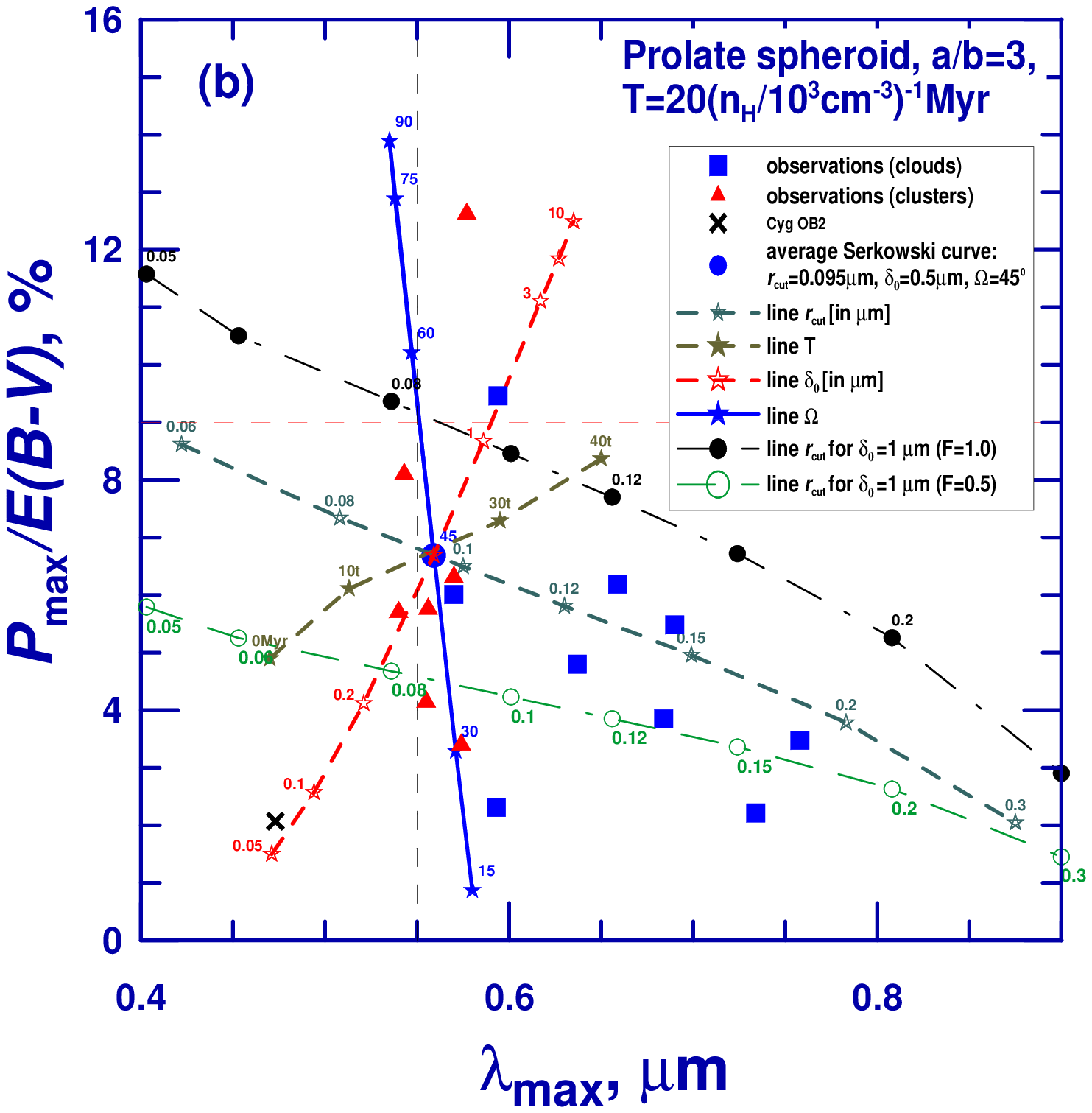}}
\caption{Polarizing efficiency $P_{\max}/E(B-V)$ versus the wavelength
of maximum polarization $\lambda_{\max}$.
Symbols  show the observational data presented in Table~\ref{t_st1}
(columns 7 and 8) and Fig.~\ref{obsd}.
Stars ($\star$) connected with solid and dashed lines
illustrate theoretical results calculated for the model:
prolate spheroids, $a/b=3$, $T=0$~Myr (a) and $T=20$~Myr (b).
Large filled circles correspond to the models
closest to the average observational Serkowski curve (see Fig.~\ref{p0}).
 Circles connected with dot-dashed lines
illustrate theoretical results calculated for the models
with $T=20$~Myr and the depolarization factors
$F=1$ (filled circles) and $F=0.5$ (open circles).
Panels illustrate the variations of the different model parameters.}
\label{pall}
\end{figure*}

If we assume that the polarizing efficiency  on average is
$P_{\max}/E(B-V) = 4 - 6$~\%, 
the average observational Serkowski curve
can be reproduced by {\rm the} models with $\Omega=60\degr$ ($T=0$~Myr) and
$\Omega=45\degr$ ($T=20$~Myr) {\rm for $F=1$}.
 These models are plotted by large filled circles in the left and right
 panels of Fig.~\ref{pall}, respectively.
 Using the circles as starting points,
we have calculated a set of models with reasonable variations
of the four mentioned parameters.
 The final picture thus  obtained resembles an {\it octopus}
shown in Fig.~\ref{pall} where just one model parameter
varies along each line.
 A distinguishing feature of the Figure is
the different behaviour of these four lines, which
allows one to estimate the effects of changes of the parameters
or their combination.  Evidently, the additional information
about the width of the polarization curve
(the parameter $K$ of the Serkowski curve) would be helpful.
Variations of the observed polarization with changes of the model
parameters show the trends summarized in
Table~\ref{t_t}. 
 For $F<1$, the octopus moves to the bottom accordingly.

\begin{table}
\bc
 \caption{{\rm Changes of the observed characteristics in dependence on the}
 model parameters.}\label{t_t}
  \begin{tabular}{ccccc} \hline
Parameter & $P_{\max}/E(B-V)$ & $\lambda_{\max}$ &  $K$ &  $R_V$\\
\hline
$r_{V,\rm cut, Si}$ \,$\nearrow$
                             & $\searrow$ & $\nearrow$   & $\nearrow$ & 0       \\
$T$ \,$\nearrow$
                             & $\nearrow$ & $\nearrow$   & $\searrow$ & $\nearrow$\\
$\delta_{0,\,{\rm Si}}^{\rm IDG}$ \,$\nearrow$
                             & $\nearrow$ & $\nearrow$   & 0          & 0         \\
$\Omega$ \,$\nearrow$
                             & $\nearrow$ & $\searrow$   & 0          & 0         \\
\noalign{\smallskip}
prolate $\rightarrow$ oblate & $\nearrow$ & $\nearrow$   & $\nearrow$ & $\searrow$\\
$a/b$ (prolate)\,$\nearrow$  & $\nearrow$ & $\searrow$   & $\searrow$ & $\searrow$\\
$a/b$ (oblate)\,$\nearrow$   & $\nearrow$ & 0            & $\nearrow$ & $\searrow$\\
\hline
\end{tabular}\ec
Note: $\nearrow$ or $\searrow$ indicate the increase or decrease of
the corresponding quantity with the growth of parameter.
0 means a little change of the quantity.\\
Parameters:
$r_{V,\rm cut, Si}$ -- minimum size of polarizing Si grains;
$T$ -- time of grain processing; 
$\delta_{0,\,{\rm Si}}^{\rm IDG}$ -- degree of grain alignment;
$\Omega$ -- direction of magnetic field.
\end{table}

 Figure~\ref{pall}a demonstrates that the models with
unprocessed dust grains ($T=0$ Myr) cannot explain the
observational points in the right and upper parts of the plots
$P_{\max}/E(B-V)$ vs $\lambda_{\max}$.
 The models with a higher value of the alignment parameter
$\delta_{0,\,{\rm Si}}^{\rm IDG}$ do not solve the problem.
Thus, we need to consider the models with larger dust grains
resulting from dust evolution in molecular clouds
(see Fig.~\ref{pall}b). As follows from this figure,
the models with $T=20$ Myr permit to reproduce the average
observational data for clouds and clusters under consideration
if we vary three additional parameters:
$r_{V,\rm cut, Si}$, $\delta_{0,\,{\rm Si}}^{\rm IDG}$ and $\Omega$.
This fact illustrate the models plotted by dot-dashed lines
in Fig.~\ref{pall}b for $\delta_{0,\,{\rm Si}}^{\rm IDG}=1.0\,\mkm$
and two values of the depolarization factor $F$.
The upper curve shows the results in the case of the regular
magnetic field ($F=1$) while the lower curve is obtained for the
model that takes into account
the turbulent component of the magnetic field ($F=0.5$).

A detailed analysis of observational data for separate objects
must be made more carefully involving information on  \is extinction
and will be carried out in the next paper.

\subsection{Variations of grain type and shape}\label{ab_v}

The \is polarization is often stated to  provide information about
 the shape of cosmic dust grains (see, e.g., \citealt{w03}).
 So far such modelling  has been made for infinite circular cylinders
 and spheroids (see \citealt{voshchinnikov12} for a review).
The infinite cylinders are the simplest
non-spherical particles whose shape cannot be varied.
 The spheroids were considered just in a few papers.
 In particular, \citet{rm79} and \citet{km95} used prolate
and oblate spheroidal grains with different aspect ratios $a/b$.
 But their consideration was restricted by perfectly aligned
rotating/non-rotating particles with a fixed orientation of the
magnetic field relative to the line of sight $90\degr$.
However, in this case the polarizing efficiency should be
in several times larger than the observed maximum  even for $F = 0.5$.

\begin{figure*}
\resizebox{8.5cm}{!}{\includegraphics{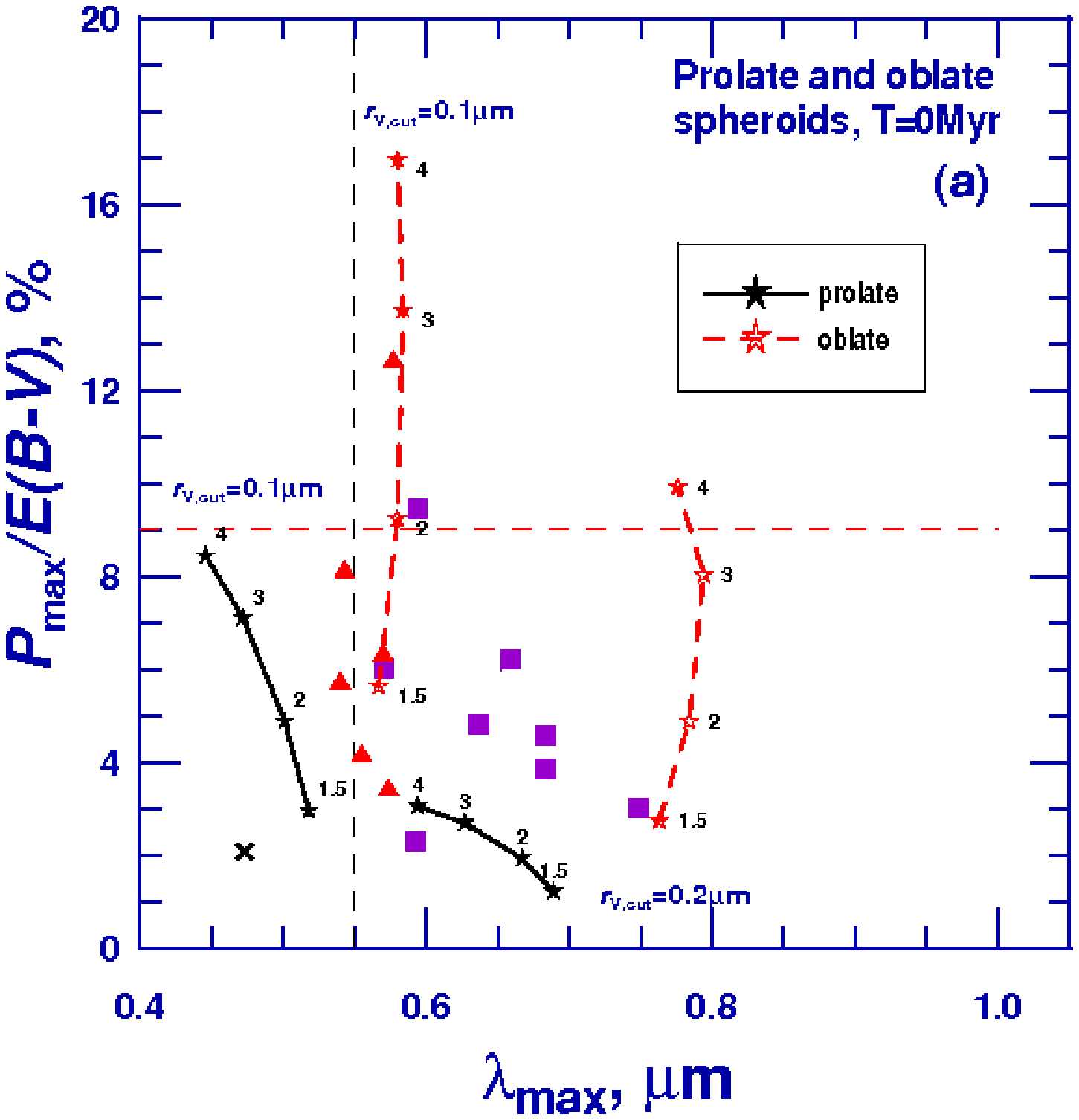}}
\resizebox{8.5cm}{!}{\includegraphics{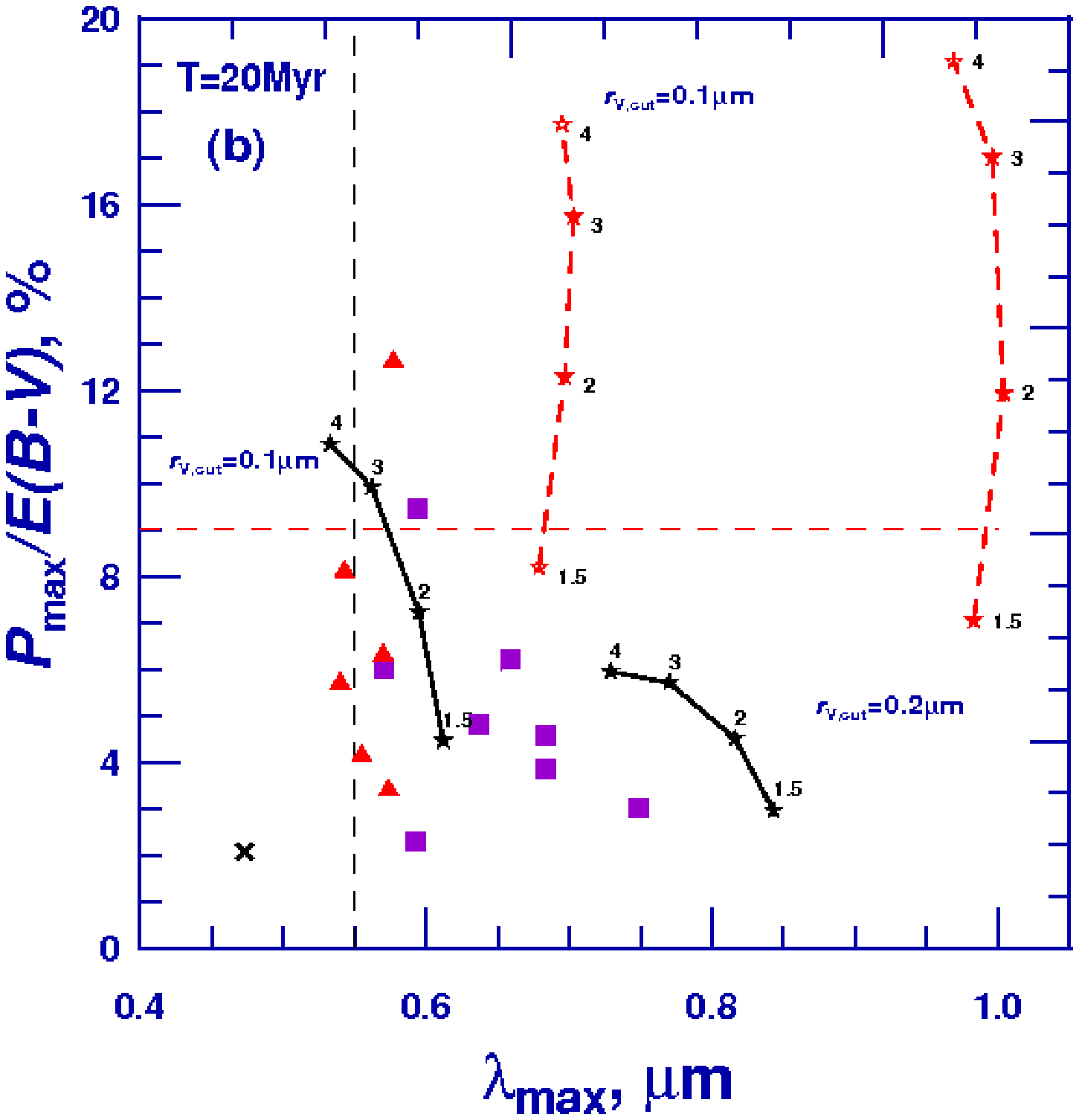}}
\caption{Polarizing efficiency $P_{\max}/E(B-V)$ versus the wavelength
of maximum polarization $\lambda_{\max}$.
Symbols  show the observational data
presented in Table~\ref{t_st1}
{\rm (columns 7 and 8)} and Fig.~\ref{obsd}.
Stars ($\star$) connected with solid and dashed lines
demonstrate theoretical results calculated for the model:
$\delta_{0,\,{\rm Si}}^{\rm IDG}=0.5\,\mkm$, $\Omega=60\degr$,
$T=0$~Myr (a) and $T=20$~Myr (b).
Panels illustrate the variations of the particle type (prolate/oblate)
and particle shape (aspect ratio $a/b$).
The values of the parameters $r_{V,\,\rm cut}$ and
$a/b$ are indicated.}
\label{p_ab}
\end{figure*}

We have calculated the \is extinction and polarization
for prolate and oblate spheroids with four aspect ratios
$a/b= 1.5, 2, 3$ and 4.  Some of these  results are given in
Fig.~\ref{p_ab} and Table~\ref{t_ab} for
$r_{V,\,\rm cut}=0.1,\, 0.2\,\mu$m and  $T=0,\,20$~Myr.
 Comparing Fig.~\ref{p_ab} with Fig.~\ref{pall}
we can arrive to the conclusion that for
prolate spheroids variations of the aspect ratio seem to be less
important than variations of other model parameters, namely
$r_{V,\rm cut, Si}$, $T$, $\delta_{0,\,{\rm Si}}^{\rm IDG}$
and $\Omega$.  When the parameter $a/b$ grows, $\lambda_{\max}$
reduces for prolate spheroids but remains almost the same
for oblate spheroids.  Note also that the oblate particles are
more efficient polarizers and can easily explain
{\rm extremely} large values of the polarizing efficiency observed,
for example, in Musca (Fig.~\ref{p_ab}a) {\rm when $F\sim0.8$}.

\begin{table*}
 \centering
\caption{{\rm Observed characteristics in dependence on the
 particle shape and time of grain processing $T$
 (the model with} $\delta_{0,\,{\rm Si}}^{\rm IDG}=0.5\,\mkm$,
 $\Omega=60\degr$).}\label{t_ab}
  \begin{tabular}{cccccccccc} \hline
&\multicolumn{4}{c}{$T=0$ Myr}&&\multicolumn{4}{c}{$T=20$ Myr} \\
  \cline{2-5}    \cline{7-10} \\
$a/b$ & $P_{\max}/E(B-V)$ & $\lambda_{\max}$ &  $K$  & $R_V$ &
      & $P_{\max}/E(B-V)$ & $\lambda_{\max}$ &  $K$  & $R_V$\\
\hline
&\multicolumn{4}{c}{Prolate spheroid, $r_{V,\rm cut}=0.1\,\mu$m}&
&\multicolumn{4}{c}{Prolate spheroid, $r_{V,\rm cut}=0.1\,\mu$m}\\
  1.5  & ~2.96   &  0.518 & 1.135 & 3.633 && ~4.46  & 0.612 & 1.013  & 6.060   \\
  2.0  & ~4.88   &  0.501 & 1.073 & 3.580 && ~7.23  & 0.595 & 0.969  & 5.971   \\
  3.0  & ~7.10   &  0.472 & 0.983 & 3.397 && ~9.91  & 0.562 & 0.904  & 5.534   \\
  4.0  & ~8.44   &  0.446 & 0.925 & 3.218 && 10.84  & 0.533 & 0.861  & 5.045   \\
\noalign{\smallskip}
&\multicolumn{4}{c}{Prolate spheroid, $r_{V,\rm cut}=0.2\,\mu$m}&
&\multicolumn{4}{c}{Prolate spheroid, $r_{V,\rm cut}=0.2\,\mu$m}\\
  1.5  & ~1.21    & 0.689  & 1.370 &&& ~2.956  & 0.843   & 1.326  & \\
  2.0  & ~1.93    & 0.667  & 1.282 &&& ~4.506  & 0.816   & 1.246  & \\
  3.0  & ~2.68    & 0.627  & 1.161 &&& ~5.708  & 0.770   & 1.136  & \\
  4.0  & ~3.05    & 0.594  & 1.088 &&& ~5.947  & 0.729   & 1.070  & \\
\noalign{\smallskip}
&\multicolumn{4}{c}{Oblate spheroid, $r_{V,\rm cut}=0.1\,\mu$m}&
&\multicolumn{4}{c}{Oblate spheroid, $r_{V,\rm cut}=0.1\,\mu$m}\\
  1.5  &  ~5.62   & 0.567  & 1.357 & 3.495 && ~8.18  & 0.679   & 1.050  & 5.708  \\
  2.0  &  ~9.23   & 0.580  & 1.442 & 3.384 && 12.29  & 0.697   & 1.107  & 5.146  \\
  3.0  &  13.72   & 0.584  & 1.524 & 3.197 && 15.73  & 0.703   & 1.194  & 4.266  \\
  4.0  &  16.95   & 0.580  & 1.681 & 3.109 && 17.72  & 0.695   & 1.306  & 3.822  \\
\noalign{\smallskip}
&\multicolumn{4}{c}{Oblate spheroid, $r_{V,\rm cut}=0.2\,\mu$m}&
&\multicolumn{4}{c}{Oblate spheroid, $r_{V,\rm cut}=0.2\,\mu$m}\\
  1.5  & ~2.73    & 0.763  & 1.700 &&& ~7.05   & 0.983  & 1.297  &  \\
  2.0  & ~4.87    & 0.784  & 1.877 &&& 11.93   & 1.004  & 1.489  &  \\
  3.0  & ~8.02    & 0.794  & 2.035 &&& 17.01   & 0.996  & 1.820  &  \\
  4.0  & ~9.91    & 0.776  & 2.242 &&& 19.07   & 0.969  & 1.992  &  \\
\hline
\end{tabular}
\end{table*}

A difference between prolate and oblate spheroids also
appears in the shape of the polarization curve:
a growth of the parameter $a/b$ makes the curve
wider ($K$ reduces) for prolate particles and
narrower ($K$ increases) for oblate particles
(Table~\ref{t_ab}).\footnote{  These tendencies
are noticed in Table~\ref{t_t}.}
VH14 compiled the parameters of Serkowski curve
for a sample of 160 lines of sight and found the observational
limits: $K \approx 0.5 -1.5$ and $\lambda_{\max} \approx 0.35 - 0.8\,\mkm$.
As follows from Table~\ref{t_ab}, the major part of the models
with oblate particles is beyond these limits.

 A more detailed consideration of the shape effects is outside
the scope of this work and requires a further investigation.

\subsection{General discussion}

It is evident that any modelling allows determining just of
intervals for  the model parameter values.
 Nevertheless, it provides some trends
in the behaviour of the observed characteristics
(Table~\ref{t_t}). Using the diagrams $P_{\max}/E(B-V)$ vs. $\lambda_{\max}$
(Figs.~\ref{p0} -- ~\ref{p_ab}) and $K$ vs. $\lambda_{\max}$ (VH14)
we can conclude that our model
has two key parameters:
 the threshold on the size of aligned silicate grains
$r_{V,\rm cut, Si}$ and time of grain processing $T$.
 After comparing with observations, it is possible to restrict
the parameters:
$0.05 \,\mkm \la r_{V,\rm cut, Si} \la 0.2 \,\mkm$ and
$T \la 20 - 30$~Myr.\footnote{This time does not contradict to the lifetime
of molecular clouds (\citealt{pag11}; \citealt{dobbs13}) because
the time-scales of accretion and coagulation
$\propto n_\mathrm{H}^{-1}$ (\citealt{hv14}), i.e. if we adopt
$n_\mathrm{H}=10^4$ cm$^{-3}$ instead of $10^3$ cm$^{-3}$,
the same size distributions are reached at one-tenth the time.}
 Note that a growth of $T$  is
accompanied by an increase of the total to selective extinction ratio $R_V$
(\citealt{hv14}; Table~\ref{t_ab}).
 Variations of the key parameters would explain the general observational
trends while additional polarization details could be reproduced
if we changed degree and direction of grain alignment
$\delta_{0,\,{\rm Si}}^{\rm IDG}$ and $\Omega$.
Calculations show that we can hardly distinguish the cases
 when the direction of the magnetic field is near the
line of sight ($\Omega \la 15\degr$) or close to perpendicular to
the line of sight ($\Omega = 75\degr - 90\degr$).
A distiction of the degree of grain alignment also presents a problem
if $\delta_{0,\,{\rm Si}}^{\rm IDG} \ga 3 - 5 \,\mkm$.

The overall picture of variations of the polarizing efficiency
presented above would not dramatically change if we
varied the aspect ratio $a/b$ of prolate grains
(Fig.~\ref{p_ab}, Table~\ref{t_ab}). 
Thus determining the shape of the particles like `cigars'
would be problematic.

The polarizing efficiency for oblate particles significantly
exceeds that of prolate ones. Thus, it is possible to
reduce the alignment degree or particle aspect ratio.
However, very flattened `pancakes' ($a/b \ga 3$) must be ruled out
because they produce too narrow polarization curves.

\section{Conclusions}\label{concl}

The main results of the paper can be formulated as follows:

\begin{enumerate}

\item
We collected the observational data about the polarizing efficiency
$P_{\max}/E(B-V)$ and the position of the maximum polarization
$\lambda_{\max}$ for  243 stars in 17 objects separated into two groups:
dark clouds and open clusters. The average data for these
two groups are distinguished by the parameter $\lambda_{\max}$.
For open clusters, the mean values of $\lambda_{\max}$
are grouped around the average \is value 0.55$\mu$m with
rather small deviations. For dark clouds,
the values of $\langle \lambda_{\max} \rangle $ may be significantly
larger than 0.55$\mu$m and have a wide scattering.

\item
We utilized the model applied earlier by \citet{vh14} and including
time evolution of the grain size distribution due to
the accretion and coagulation  processes and
new optical constants of grain materials.
 To calculate the polarization curves,
we used homogeneous silicate and carbonaceous spheroidal
particles of different aspect ratios $a/b$ having  imperfect alignment.
 It was assumed that polarization was mainly produced by
large silicate particles
with sizes $r_{V} \ga r_{V,\rm cut}$.
 We calculated the wavelength dependence of extinction and polarization
and determined the parameters $P_{\max}/E(B-V)$ and $\lambda_{\max}$.

\item
We focused on the interpretation of the observed
relation between $P_{\max}/E(B-V)$ and $\lambda_{\max}$ keeping
in mind probable depolarization effect caused by the random component
of the magnetic field.
 Theoretically both quantities are mainly determined by
two key parameters: the threshold on the size of aligned silicate grains
$r_{V,\rm cut}$ and time of grain processing $T$.
 We found that the models with the initial size distribution (no grain
processing, $T=0$~Myr) {\rm that reproduce the average curve of
the interstellar extinction fail} to explain the data with
$\lambda_{\max} \ga 0.65\,\mkm$ observed in several dark clouds.
 An inclusion of evolutionary effects allows us to find
the models which give the polarizing efficiency
and $\lambda_{\max}$ values similar to the average observed ones
  for reasonable depolarization ($F \sim 0.5-0.8$).
This occurs for $0.05 \,\mkm \la r_{V,\rm cut, Si} \la 0.2 \,\mkm$ and
$T \la (20 - 30)(n_\mathrm{H}/10^3 \mathrm{cm}^{-3})^{-1}$~Myr
and is in agreement with the lifetime of molecular clouds.
The additional polarization details are reproduced
if we change the degree ($\delta_{0}^{\rm IDG}$) and
direction ($\Omega$) of particle orientation.
 However, these results cannot yet be a conclusive evidence
that the dust evolution only can completely explain 
the observational data under consideration
because of a rather large number of the parameters still involved
in our modeling.  A more detailed analysis in particular that
involving the width of the polarization curve is required.

\item
We found that  generally the change of the aspect ratio $a/b$ of
prolate grains would not dramatically influence the variations of
the polarizing efficiency that makes the determination of
the grain shape rather problematic for prolate particles.
 The polarizing efficiency for oblate spheroids significantly
exceeds that of prolate ones but particles with $a/b \ga 3$
must be excluded because they produce too narrow polarization curves
that are not observed.
\end{enumerate}

\section*{Acknowledgments}

We are grateful to anonymous referee for very useful comments
and suggestions and thank A. P. Jones for sending us
the refractive indexes in the tabular form
and interesting discussions.
 We acknowledge the support from RFBR grant 16-02-00194
and RFBR--DST grant 16-52-45005.


\bsp
\label{lastpage}
\end{document}